\documentclass[lettersize,journal]{IEEEtran}
\usepackage{amsmath,amsfonts}
\usepackage{algorithmic}
\usepackage{algorithm}
\usepackage{array}
\usepackage[caption=false,font=footnotesize,labelfont=sf,textfont=sf]{subfig}
\usepackage{textcomp}
\usepackage{stfloats}
\usepackage{url}
\usepackage{verbatim}
\usepackage{graphicx}
\usepackage{cite}

\usepackage{caption}
\usepackage{amsfonts,amssymb}
\usepackage{bm}
\usepackage{graphicx}
\usepackage{booktabs}
\usepackage{epstopdf}
\usepackage{ragged2e}
\usepackage{tablefootnote}
\usepackage{color}
\begin{document}

\title{Fairness Optimization for Intelligent Reflecting Surface Aided Uplink Rate-Splitting Multiple Access}

\author{Shanshan~Zhang,
	Wen~Chen,~\IEEEmembership{Senior Member,~IEEE,}
	Qingqing~Wu,~\IEEEmembership{Senior Member,~IEEE,}
	Ziwei~Liu,
	Shunqing~Zhang,~\IEEEmembership{Senior Member,~IEEE,}
	Jun~Li,~\IEEEmembership{Senior Member,~IEEE}
	\thanks{
		Part of this article was accepted by 2023 IEEE Global Communications Conference (Globecom 2023) \cite{RN533}.
		
		Shanshan Zhang, Wen Chen, Qingqing Wu, and Ziwei Liu are with the Department of Electronic Engineering, Shanghai Jiao Tong University, Shanghai 200240, China (e-mail: shansz@sjtu.edu.cn;	wenchen@sjtu.edu.cn; wu.qq1010@gmail.com; ziweiliu@sjtu.edu.cn).
		
		Shunqing Zhang is with the Shanghai Institute for Advanced Communication and Data Science, Shanghai University, Shanghai 200444, China (e-mail:shunqing@shu.edu.cnn).
		
		Jun Li is with the School of Electronic and Optical Engineering,
		Nanjing University of Science and Technology, Nanjing 210094, China
		(e-mail: jun.li@njust.edu.cn).
	}
}

\maketitle

\begin{abstract}
This paper studies the fair transmission design for an intelligent reflecting surface (IRS) aided rate-splitting multiple access (RSMA). IRS is used to establish a good signal propagation environment and enhance the RSMA transmission performance. The fair rate adaption problem is constructed as a max-min optimization problem. To solve the optimization problem, we adopt an alternative optimization (AO) algorithm to optimize the power allocation, beamforming, and decoding order, respectively. A generalized power iteration (GPI) method is proposed to optimize the receive beamforming, which can improve the minimum rate of devices and reduce the optimization complexity. At the base station (BS), a successive group decoding (SGD) algorithm is proposed to tackle the uplink signal estimation, which trades off the fairness and complexity of decoding. 
At the same time, we also consider robust communication with imperfect channel state information at the transmitter (CSIT), which studies robust optimization by using lower bound expressions on the expected data rates. Extensive numerical results show that the proposed optimization algorithm can significantly improve the performance of fairness. It also provides reliable results for uplink communication with imperfect CSIT.
\end{abstract}


\section{Introduction}
\IEEEPARstart{I}{n} the sixth-generation mobile communications (6G), the application of the Internet of Things (IoT) in various fields will spur explosive growth in the number of IoT devices. The evolution of IoT to the Internet-of-Everything (IoE) will speed up. Wireless systems need to support more devices and provide services with higher throughput, ultra-reliability, and heterogeneous quality of service (QoS). Therefore, how to support more devices and provide a high data rate with high reliability is becoming a new challenge. Non-orthogonal multiple access (NOMA) is proposed to combine with grant-free protocols to meet the requirements of massive access\cite{RN477}. However, NOMA can be vulnerable to the channel state information at the transmitter (CSIT), which will result in reduced communication rates for partial devices. To address these issues, rate-splitting multiple access (RSMA) has been proposed as a design of both the physical (PHY) layer and the multiple access technique \cite{RN450}. 

For uplink communication, RSMA is more general than NOMA and orthogonal multiple access (OMA) \cite{RN497}. It splits the signal into two sub-messages at the transmitter. Then, the two sub-messages are independently encoded and allocated two different powers. Finally, each device superposes the encoded sub-messages and transmits them to the base station (BS). Uplink RSMA reduces to uplink NOMA if the transmit power of each device is fully allocated to one message \cite{9831440}. For downlink RSMA, the message intended for each device is split into a common and a private message at the BS. After jointly encoding the common parts into a common message for decoding by all devices and independently encoding the private parts into private messages for the corresponding devices, the BS linearly precodes all encoded messages and broadcasts the superimposed messages to all devices.  Different from the existing schemes that fully treat interference as noise or fully decode interference, RSMA partially decodes the interference from other messages and partially treats them as noise at the receiver. In this manner, RSMA offers improved interference management. It provides a new paradigm for massive connectivity, which bridges the two extremes of fully decoding interference and fully treating interference as noise \cite{9831440}. In comparison to NOMA, RSMA derives its advantages from two key factors. Firstly, the number of decoding orders in RSMA is larger than that in NOMA, which shows that RSMA is a robust scheme. Secondly, the rate allocation for each device in RSMA consists of a sum of two sub-messages, whereas in NOMA, it involves decoding a single message only.  Consequently, RSMA has been recognized as a promising scheme for non-orthogonal transmission, interference management, and the implementation of massive access strategies in the context of 6G.
	
Additionally, RSMA can enhance spectral efficiency through the aid of Intelligent Reflecting Surface (IRS). IRS technology is employed to enhance network coverage and resolve blockage issues in wireless communications \cite{RN473, RN475, RN457, RN458, RN479}. As a kind type of metasurface, IRS is able to change the end-to-end signal propagation direction with low-cost passive components \cite{RN416}. Leveraging IRS, wireless propagation environments can be controlled through a cost-effective scheme while avoiding the deployment of additional power-hungry and expensive communication schemes. Therefore, IRS has attracted significant attention and research interest in the field of wireless communications. 
\subsection{Related works}
The terminology ``RSMA'' is mentioned in \cite{RN365} for multiple access channels (MAC). The uplink RSMA is proven to achieve the capacity region of the $K$-device Gaussian MAC without time-sharing \cite{RN365, RN372, RN370}. RSMA is introduced for multiple-input multiple-output (MIMO) wireless networks in various scenarios such as MU-MIMO, massive MIMO, and multi-cell coordination \cite{RN338}. Kinds of literature \cite{RN339, RN449} generalize RSMA to massive MIMO deployments with imperfect CSIT, and study the performance of RSMA. Rate-splitting is also applied to achieve max-min fairness amongst multiple co-channel multicast groups through transmit beamforming \cite{RN357}. \cite{RN454} and \cite{RN451} try to enhance the max-min spectral efficiency by precoder designs in RSMA-aided downlink transmission. \cite{RN480} investigates the spectral and energy efficiency tradeoff of RSMA in multi-user multi-antenna systems.

IRS leverages reconfigurable reflections, providing significant advantages over conventional relaying protocols. It enables a flexible precoding design and enhances the resilience of RSMA to imperfect successive interference cancellation (SIC). Furthermore, RSMA contributes to robust IRS optimizations even under imperfect CSIT \cite{RN355}. An IRS-RSMA architecture is proposed in \cite{RN453}, where a closed-form expression for outage probability is derived. This architecture demonstrates substantial performance gains compared to the IRS-aided NOMA framework when utilizing a single-antenna BS. Additionally, \cite{RN426} studies the sum-rate maximization for considered uplink RSMA system, focusing on joint optimization of power allocation to uplink devices and beamforming design. \cite{RN407} proposes an IRS-aided RSMA system to improve outage performance and guarantee QoS requirements. 
IRS-aided RSMA systems are further explored in \cite{RN450} to enhance fairness and in \cite{RN478} to maximize the minimum secrecy rate in multi-antenna broadcast channels.
\cite{RN474} jointly optimizes parameters of IRS and RSMA to improve energy efficiency and spectral efficiency. Moreover, \cite{RN534} investigates the weighted sum-rate maximization problem by jointly optimizing the power allocation, IRS transmissive coefficients, and common rate allocation. To enhance robustness against imperfect CSIT, A deep learning (DL) based RSMA is proposed in \cite{RN452}. 

There are several works \cite{RN368,RN367,RN426,RN407,RN481} that investigate uplink communications. Uplink RSMA systems are proposed to enhance outage performance \cite{RN407} and fairness \cite{RN368} in a two-device MAC. \cite{RN367} focuses on the power allocation in the transmitter and decoding order at the BS to maximize the sum-rate. Furthermore, \cite{RN481} studies the impact of block length and target rate on the throughput and error probability performance of uplink RSMA. 

\subsection{Motivations}
Most existing works focus on downlink RSMA systems rather than uplink RSMA systems.  Although several works study the uplink RSMA, they typically only consider the two-device MAC.  However, uplink RSMA can achieve the capacity region of MAC and enhance fairness. Therefore, the potential value of uplink RSMA remains to be exploited. Moreover, the performance of RSMA is strongly dependent on the SIC \cite{RN365, RN338, RN426, RN367}. But in the uplink system with massive connectivity, RSMA encounters challenges of complexity issues and SIC processing delay. In an uplink RSMA system, the complexity of SIC is directly proportional to the number of sub-messages. With each additional sub-message, an extra decoding and interference cancellation process is required. Specifically, in a system with $K$ devices, the receiver needs $2K$ layers of SIC to decode all $2K$ sub-messages.
 Therefore, implementing RSMA in uplink wireless networks also faces several challenges, including designing efficient decoding schemes and managing resources for signal transmission.

In order to reduce the complexity and the time delay of the signal processing, we apply successive group decoding (SGD) in the uplink RSMA system, which is introduced in \cite{RN438}. SGD is an extension of the conventional SIC. Unlike SIC, SGD allows for the joint decoding of a subset of devices at each decoding stage, rather than just one. As a result, SGD can reduce the number of decoding layers compared to SIC. Therefore, SGD can reduce the complexity of decoding at the BS and take advantage of RSMA to enhance fairness.

Some works design IRS-aided RSMA systems to achieve better rate performance and enhance coverage capability for multiple devices \cite{RN426, RN452, RN450}. However, the propagation environment may be complicated in practice. To deploy IRS, an exhaustive search for the optimal location requires the global CSIT at all locations, which is practically difficult to obtain \cite{RN457,RN538}. As a flexible interference management framework, the RSMA technique can achieve robustness to imperfect CSIT \cite{RN355}. Thus, the IRS is suitable to assist RSMA systems with imperfect CSIT.
An important consideration arises regarding how to satisfy QoS requirements in IRS-aided uplink RSMA systems with imperfect CSIT. Addressing this issue is crucial for ensuring reliable and efficient communication in such systems.

\subsection{Contributions}

In this paper, we study an IRS-aided uplink RSMA system for the massive MIMO system. The main contributions of this work are summarized as follows:
\begin{itemize}
	\item In IRS-aided uplink communication, multiple devices can share the same resource blocks at the same time. We consider the direct channel and the IRS-aided channel to improve channel gain. By formulating a max-min fairness optimization problem, we jointly optimize the design of power allocation in the transmitter, receive beamforming at the BS, phase-shift beamforming at IRS, and decoding order. Subsequently, we adopt an alternative optimization (AO) algorithm to iteratively optimize the power allocation, receive beamforming, phase-shift beamforming, and decoding order.
	\item At the receiver, we adopt the SGD, which decodes signals using linear detection within each group and applies SIC between groups. Group decoding order plays an essential role in SGD since different group decoding orders will result in different decoding outputs and further different system throughputs \cite{RN405}. Therefore, we propose a greedy grouping algorithm with low complexity to design the group decoding order and achieve fairness.
	\item We employ a LogSumExp approximation to develop a generalized power iteration (GPI) method for obtaining the optimal solution of receive beamforming at the BS. Compared to conventional methods based on a convex semidefinite program (SDP), GPI offers lower complexity and superior performance.
	\item For IRS-aided uplink RSMA with imperfect CSIT, we derive a lower bound on device rates. We then aim to maximize the lower bound of the minimum device rate by jointly optimizing the design of power allocation, group decoding order, receive beamforming, and phase-shift beamforming.
	
\end{itemize}

\subsection{Organization}
The rest of this paper is organized as follows. Section \ref{SM} describes the system model and introduces the SGD at the receiver. Section \ref{CSI} formulates the fair rate adaption as a max-min problem for the uplink RSMA system with known CSIT and adopts an AO algorithm to iteratively optimize the power allocation, receive beamforming, phase-shift beamforming, and decoding order.
Section \ref{imCSI} discusses the max-min fairness optimization problem with imperfect CSIT. Section \ref{Nu} analyzes the numerical results obtained from the proposed scheme. Finally, Section \ref{Con} concludes.
\subsection{Notation}

Throughout this paper, scalars, vectors, and matrices are denoted by the italic lowercases (e.g., $x$), bold italic lowercases (e.g., $\bm x$), and bold uppercase (e.g., $\mathbf X$), respectively. Let $\mathbf I$ denote the unit matrix. $\bm 0_N$ denotes a column vector with $N$ zero elements. Use calligraphy uppercases (e.g., $\mathcal{N})$ to represent sets. $|\mathcal{N}|$ is the number of elements in set $\mathcal{N}$. $x_{ij}=[\mathbf X]_{ij}$ denotes the $(i, j)$-th element of matrix $\mathbf X$. $[\mathbf X]_{i:}$ denotes the $i$th row of matrix $\mathbf X$. $[\bm x]_{i}$ denotes the $i$th element of vector $\bm x$. 
The transpose, complex conjugate, and conjugate transpose operators are denoted by $(\cdot)^T$, $(\cdot)^{\dagger}$, and $(\cdot)^H$, respectively. $\text{tr}(\cdot)$ denotes the trace part of the term.  $\|\cdot\|_2$ denotes the 2-norm.

\section{System Descriptions}
\label{SM}
In this section, we first introduce the IRS-aided uplink RSMA system and then present the SGD scheme. 

\subsection{System Model}

We consider an uplink system, which consists of $K$ single-antenna devices, a BS equipped with $M$ antennas, and an IRS composed of $N$ elements. In the RSMA system, the $K$ original messages coming from $K$ devices are split into $2K$ sub-messages. Denote $x_{k,i}$ as the $i$th sub-message of device $k$. Accordingly, power constraints are assigned to these sub-messages to satisfy the original constraints. 
$p_{k,i}$ denotes the power allocation for the $i$th sub-message of device $k$, where $i=1,2$. Each device $k$ has a maximum transmit power limit $P_{\text{max}}$, i.e.,$\sum\limits_{i=1}^2 p_{k,i}\leq P_{\text{max}}, 1\le k\le K$.
At the BS, the received signal $\bm y\in\mathbb{C}^{M\times 1}$ is 
\begin{eqnarray}	
\bm y= \sum\limits_{k=1}^K\left(\mathbf H_{rb}\mathbf\Theta\bm h_{sr,k}+\bm h_{d,k}\right)\left(\sqrt{p_{k,1}}x_{k,1}+\sqrt{p_{k,2}}x_{k,2}\right)+\bm w,\nonumber
\end{eqnarray}
where $\mathbf H_{rb}\in \mathbb{C}^{M\times N}$, $\bm h_{sr,k}\in \mathbb{C}^{N\times 1}$, and $\bm h_{d,k}\in \mathbb{C}^{M\times 1}$ are the channels from IRS to the BS, from device $k$ to the IRS, and from device $k$ to the BS, respectively. $\mathbf\Theta=\text{diag} [e^{j\theta_1},\ldots,e^{j\theta_N}]$ is the phase-shift matrix, where $\theta_n\in(-\pi,\pi]$ is the phase shift induced by the $n$th element of the IRS. $\bm w\sim \mathcal{CN}(\mathbf 0,\sigma^2\mathbf I)$ is the additive white Gaussian noise (AWGN) at the BS.
The massive MIMO system follows a block-fading model where channels follow independent quasi-static flat-fading in each block of coherence time. According to \cite{RN171,RN115}, the channel matrix $\mathbf H_{rb}$ is given by 
\begin{eqnarray}
\mathbf H_{rb} = \sum\limits_{p=1}^{N_{rb}}\beta_{p}^{rb}\bm a_{B}(\theta_{B,p}^{rb})\bm a_{R}^H(\theta_{R,p}^{rb})e^{-j2\pi\tau_{p}^{rb} \frac{B_s}{2}},
\label{eq:2}
\end{eqnarray}
where $B_s$ represents the two-sided bandwidth, and $N_{rb}$ denotes the number of multi-path components (MPCs). $\beta_{p}^{rb}$ and $\tau_{p}^{rb}$ are the complex path gain and the path delay of the $p$th MPC, respectively. The array steering and response vectors are given by
\begin{eqnarray}
\begin{aligned}
\bm a_{B}(\theta_{B,p}^{rb})=& [1, e^{-j2\pi\theta_{B,p}^{rb}},\ldots,e^{-j2\pi(M-1)\theta_{B,p}^{rb}}]^T,\\
\bm a_{R}(\theta_{R,p}^{rb})=& [1, e^{-j2\pi\theta_{R,p}^{rb}},\ldots,e^{-j2\pi(N-1)\theta_{R,p}^{rb}}]^T.
\label{eq:3}
\end{aligned}
\end{eqnarray}
In (\ref{eq:3}), $\theta_{\cdot,p}^{rb}=d\sin(\phi_{\cdot,p}^{rb})/\lambda$, where $\phi_{\cdot,p}^{rb}\in [-\pi/2,\pi/2]$ is the physical angle, $\lambda$ is the wavelength of propagation, and $d=\lambda/2$ is the normalized antenna spacing. Similarly, $\bm h_{sr,k}$ and $\bm h_{d,k}$ are given by 
\begin{eqnarray}
\bm h_{sr,k} = &\sum\limits_{p=1}^{N_{sr,k}}\beta_{p,k}^{sr}\bm a_{R}(\theta_{R,p,k}^{sr})e^{-j2\pi\tau_{p,k}^{sr}\frac{B_s}{2}},\\
\bm h_{d,k} = &\sum\limits_{p=1}^{N_{d,k}}\beta_{p,k}^{d}\bm a_{B}(\theta_{B,p,k}^{d})e^{-j2\pi\tau_{p,k}^{d}\frac{B_s}{2}},
\label{eq:4}
\end{eqnarray}
where $N_{sr,k}$ and $N_{d,k}$ denote the number of MPCs for channel from device $k$ to the IRS, and from device $k$ to the BS. $\beta_{p,k}^{sr}$, $\beta_{p,k}^{d}$, $\tau_{p,k}^{sr}$, and $\tau_{p,k}^{d}$ are the complex path gain and the path delay of the $p$th MPC for channel from device $k$ to the IRS, and from device $k$ to the BS, respectively. 
Define
\begin{eqnarray}
\begin{aligned}	
\bm h_k&=\mathbf H_{rb}\mathbf\Theta\bm h_{sr,k}+\bm h_{d,k}\\&=\mathbf H_{rb}\text{diag}(\bm h_{sr,k})\text{diag}(\mathbf\Theta)+\bm h_{d,k}\\&=\mathbf H_k\bm v,
\end{aligned}
\end{eqnarray}
where $\mathbf H_k=[\mathbf H_{rb}\text{diag}(\bm h_{sr,k}), \bm h_{d,k}]\in\mathbb C^{M\times(N+1)}$ and $\bm v=[\text{diag}(\mathbf\Theta);1]=[e^{j\theta_1},\ldots,e^{j\theta_N},1]^T$.
Finally, the received signal is written as 
\begin{eqnarray}	
\bm y=\sum\limits_{k=1}^K\mathbf H_k\bm v\left(\sqrt{p_{k,1}}x_{k,1}+\sqrt{p_{k,2}}x_{k,2}\right)+\bm w.
\end{eqnarray} 
\begin{figure*}[t]
	\centering
	\begin{center}
		\includegraphics[scale=0.4]{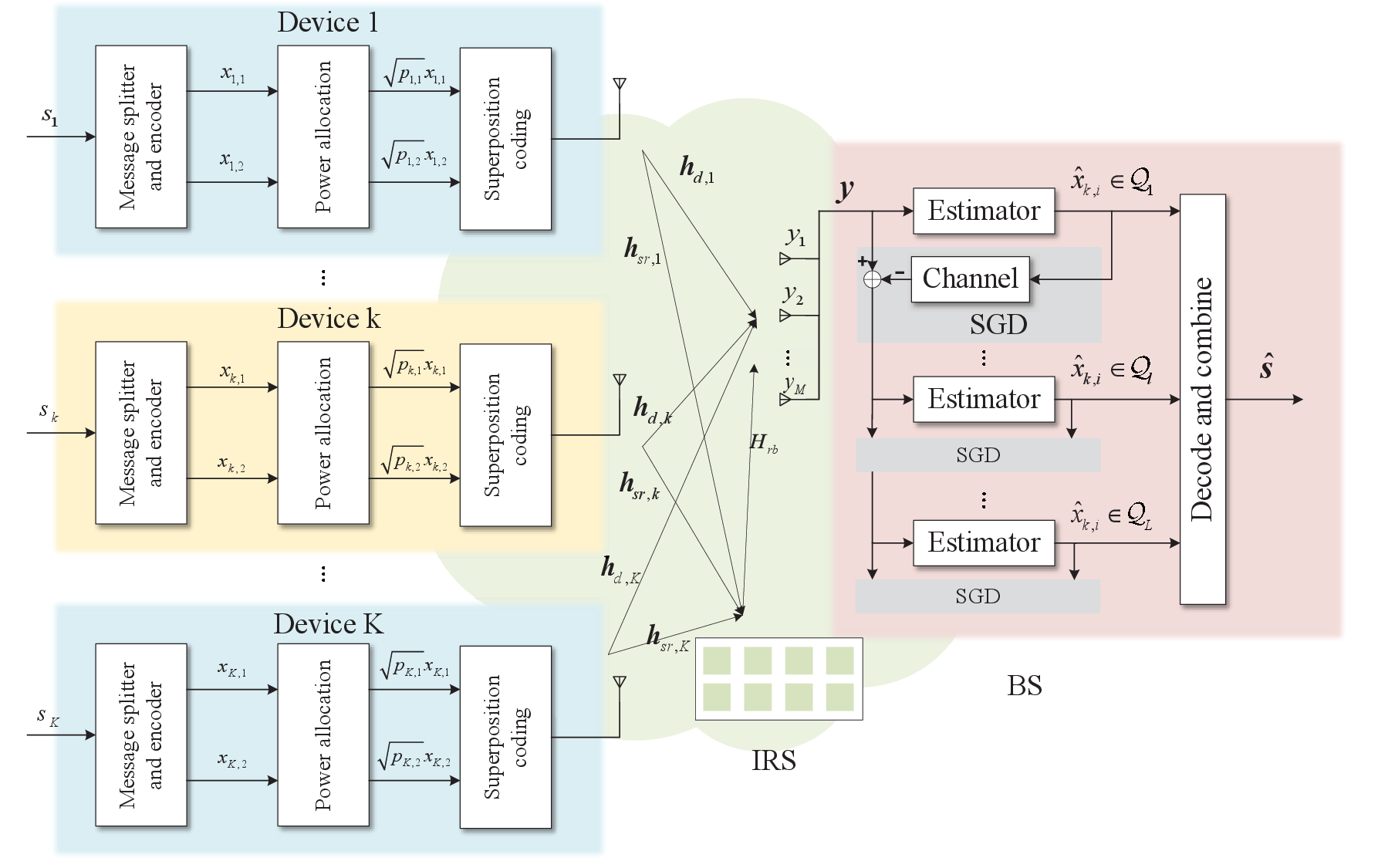}
	\end{center}
	\caption{The architecture of uplink RSMA with the SGD receiver.}
	\label{system}
	\vspace{-0.2cm}
\end{figure*}
\subsection{Successive Group Decoding (SGD)}
\label{SGD}
SGD employs linear decoding of multiple sub-messages at each stage. Assume that the BS divides the devices' messages into $L$ groups. Define set $\mathcal Q\triangleq\{(k,i)\}_{k=1,i=1}^{K,I}$ and $\mathcal Q_l\triangleq\{(k,i)|(k,i)\in\mathcal Q\}, l=1,\ldots, L$, where $(k,i)$ denotes the index of the $i$th sub-message of device $k$. There are $\mathcal Q_1\cup\cdots\cup\mathcal Q_L=\mathcal Q$ and $\mathcal Q_1\cap\cdots\cap\mathcal Q_L=\emptyset$. In the $l$th stage of the SGD scheme, sub-messages in group $\mathcal Q_l$ are decoded by treating all undecoded sub-messages as Gaussian interference. The decoding order at the BS is  $\mathcal Q_1,\ldots,\mathcal Q_L$.
In each stage, linear beamforming is employed in the received signal $\bm y$ for decoding $(k, i)$-th sub-message. Define $\underline{\mathcal Q}_l=\mathcal Q_l\cup\ldots\cup\mathcal Q_L$. If $(k,i)\in\mathcal Q_l$, the estimated signal $\hat{x}_{k,i}$ is
\begin{eqnarray}	
\begin{aligned}	
\hat{x}_{k,i}=\bm g_{k,i}^H \bm y=&\bm g_{k,i}^H\mathbf H_k\bm v \sqrt{p_{k,i}} x_{k,i}\\&+\bm g_{k,i}^H\sum\limits_{\substack{(n,j)\in\underline{\mathcal Q}_l,\\(n,j)\neq (k,i) }}\mathbf H_n\bm v\sqrt{p_{n,j}} x_{n,j} +\bm g_{k,i}^H\bm w,
\label{eq:x}
\end{aligned}	
\end{eqnarray} 
where $\bm g_{k,i}\in\mathbb C^{M\times 1}$ denotes the beamforming vector for the $i$th sub-message of device $k$. The SGD operates as follows:
\begin{itemize}
	\item[a)] Initialize with inputs: $l=1,\mathbf H_1,\ldots,\mathbf H_K,$ and $\mathcal Q_1,\ldots,\mathcal Q_L$.
	\item[b)] For $(k,i)\in \mathcal Q_l$, estimate $x_{k,j}$ according to (\ref{eq:x}).
	\item[c)] Update $\bm y = \bm y - \sum\limits_{(k,i)\in\mathcal Q_l}\mathbf H_k\bm v\left(\sqrt{p_{k,i}}x_{k,i}\right)$ and $l=l+1$.
	\item[d)] If $l=L+1$, stop, otherwise go to step b).
\end{itemize}
The uplink RSMA system is shown in Fig. \ref{system}. For $(k,i)\in \mathcal Q_l$, the rate is
\begin{eqnarray}	
r_{k,i}=\log_2\left(1+\frac{p_{k,i}\|\bm g_{k,i}^H\mathbf H_k\bm v\|_2^2}{\sum\limits_{\substack{(n,j)\in\underline{\mathcal Q}_l,\\(n,j)\neq (k,i) }}p_{n,j}\|\bm g_{k,i}^H\mathbf H_n\bm v\|_2^2+\|\bm g_{k,i}\|_2^2\sigma^2}\right).\nonumber
\label{eq:sr}
\end{eqnarray}

\section{Resource Allocation for Fairness with known CSIT}
\label{CSI}
This section aims to maximize the minimum rate of the IRS-aided uplink network with known CSIT. We formulate a max-min problem and decouple it into four sub-problems: receive beamforming optimization, phase-shift beamforming optimization, power allocation, and decoding order optimization. Then the sub-problems are optimized alternately until convergence is reached.

\subsection{Problem Formulation}
\label{P1}
Under the joint consideration of grouping order, power allocation, and beamforming design (including receive beamforming at the BS and phase-shift beamforming at IRS), we formulate the fair rate adaption problem as follows
\begin{eqnarray}
\begin{aligned}
\max_{\mathcal Q_l,\bm v ,\bm g_{k,i},p_{k,i}}& \quad \min_{k\in\mathcal K} \sum_{i=1}^2 r_{k,i}\\
\text{s.t.} & \quad \sum\limits_{i=1}^2 \|\bm g_{k,i}\|_2^2\leq P_{\text{max}}^{\text{b}}, \forall k\in\mathcal K, \\&\quad |[\bm v]_n| =1, n=1,\ldots,N,[\bm v]_{N+1}=1,\\&\quad \sum\limits_{i=1}^2 p_{k,i}\leq P_{\text{max}},\forall k\in\mathcal K,\\&\quad  p_{k,i}\ge 0,\forall (k,i)\in \mathcal Q,\label{eq:p1con2}
\end{aligned}
\end{eqnarray}
where $\mathcal K$ is the set of devices and $P_{\text{max}}^{\text{b}}$ is the maximum power limit of receive beamforming. By defining  $\mathbf V=\bm v\bm v^H$, where $\mathbf V\succeq\mathbf 0, \text{rank}(\mathbf V)=1$. We can rewrite the rate as
\begin{eqnarray}	
r_{k,i}=\log_2\left(\frac{\sum\limits_{\substack{(n,j)\in\underline{\mathcal Q}_l}}p_{n,j}\bm g_{k,i}^H\mathbf H_n\mathbf V\mathbf H_n^H\bm g_{k,i}+\sigma^2\bm g_{k,i}^H\bm g_{k,i}}{\sum\limits_{\substack{(n,j)\in\underline{\mathcal Q}_l,\\(n,j)\neq (k,i) }}p_{n,j}\bm g_{k,i}^H\mathbf H_n\mathbf V\mathbf H_n^H\bm g_{k,i}+\sigma^2\bm g_{k,i}^H\bm g_{k,i}}\right).
\label{eq:r1}\nonumber
\end{eqnarray}
The optimization problem for resource allocation design can be formulated as
\begin{eqnarray}
\begin{aligned}
\text{(P0)}\quad \max_{\mathcal Q_l,\mathbf V ,\bm g_{k,i},p_{k,i}}& \quad \min_{k\in\mathcal K} \sum_{i=1}^2 r_{k,i}\\
\text{s.t.} \quad\text{(C1)}&\quad \sum\limits_{i=1}^2 \|\bm g_{k,i}\|_2^2\leq P_{\text{max}}^{\text{b}}, \forall k\in\mathcal K, 
\\\text{(C2)}&\quad [\mathbf V]_{nn}=1, n=1,\ldots,N+1,
\\\text{(C3)}&\quad \mathbf V\succeq\mathbf 0, \text{rank}(\mathbf V)= 1,
\\\text{(C4)}&\quad \sum\limits_{i=1}^2 p_{k,i}\leq P_{\text{max}}, \forall k\in\mathcal K,
\\\text{(C5)}&\quad p_{k,i}\ge 0,\forall (k,i)\in\mathcal Q,
\end{aligned}\label{eq:p01}
\end{eqnarray}
where (C1) is the power constraint of receive beamforming. (C2) ensures the unit-modulus constraints on the phase shifts. (C3) imposes semidefinite and nonnegativity constraints of $\mathbf V$. (C4) and (C5) specify the transmit power constraints. It is obvious that (P0) is an intractable non-convex problem due to the coupled optimization variables in the objective function and the non-convex rank constraint in (C3). Therefore, this non-convex problem is hard to solve directly. In the next section, we discuss the AO algorithm to solve this optimization problem, which is widely used and empirically efficient for driving the non-convex problem with coupled optimization variables.

\subsection{Optimizing Receive Beamforming }
\label{opt_G}
We propose a GPI method to optimize receive beamforming for given phase-shift beamforming, power allocation, and decoding order. Compared to a convex SDP \cite{RN533},
the proposed algorithm does not rely on CVX and reduces computational complexity. Next, we will study how to apply the GPI method to solve the receive beamforming optimization.

Given phase-shift beamforming $\mathbf V$, power allocation $p_{k,i}$ for $(k,i)\in \mathcal Q$, and decoding order $\mathcal Q_l$ with $0\le l\le L$, (P0) is transformed into 
\begin{eqnarray}
\begin{aligned}
\text{(P1)}\quad \max_{\bm g_{k,i}}& \quad \min_{k\in\mathcal K} \sum_{i=1}^2 r_{k,i}\\
\text{s.t.} &\quad\text{(C1)}.
\end{aligned}\label{eq:p11}
\end{eqnarray}
According to \cite{RN486}, the max-min function can be approximated as a smooth function by using the LogSumExp technique. When $\alpha\to 0$, there is 
\begin{eqnarray}
\begin{aligned}
\min_{k\in\mathcal K} \sum\limits_{i=1}^2 r_{k,i}\approx -\alpha \log\left(\frac{1}{K}\sum\limits_{k\in\mathcal K}\exp\left(\frac{ \sum\limits_{i=1}^2 r_{k,i}}{-\alpha}\right)\right).
\end{aligned}
\end{eqnarray} 
Define
\begin{eqnarray}	
\begin{aligned}
\mathbf {\Gamma}_{k,i}^u&=\sum\limits_{(n,j)\in\underline{\mathcal Q}_l}p_{n,j}{\mathbf H}_n\mathbf V{\mathbf H}_n^H+\sigma^2\mathbf I,\\
\mathbf {\Gamma}_{k,i}^d&=\sum\limits_{\substack{(n,j)\in\underline{\mathcal Q}_l,\\(n,j)\neq (k,i) }}p_{n,j}{\mathbf H}_n{\mathbf V}{\mathbf H}_n^H+\sigma^2\mathbf I.
\end{aligned}
\end{eqnarray}
Then we have 
\begin{eqnarray}
\begin{aligned}
r_{k,i}=\log_2\left(\frac{\bm g_{k,i}^H\mathbf {\Gamma}_{k,i}^u\bm g_{k,i}}{\bm g_{k,i}^H\mathbf {\Gamma}_{k,i}^d\bm g_{k,i}}\right).
\end{aligned}
\end{eqnarray}
Therefore, (P1) can turn to
\begin{eqnarray}
\begin{aligned}
\text{(P1.1)}\quad \max_{\bm g_{k,i}}& \quad -\alpha \log\left(\frac{1}{K}\sum\limits_{k\in\mathcal K}\right.\\& \left.\exp\left( \sum\limits_{i=1}^2\log_2\left(\frac{\bm g_{k,i}^H\mathbf {\Gamma}_{k,i}^u\bm g_{k,i}}{\bm g_{k,i}^H\mathbf {\Gamma}_{k,i}^d\bm g_{k,i}}\right)^{-\frac{1}{\alpha}}\right)\right)\\
\text{s.t.} & \quad \text{(C1)}.
\end{aligned}\label{eq:p111}
\end{eqnarray}
For (P1.1), there is a first-order optimality condition shown in the following Lemma 1.

\textbf{Lemma 1:} The first-order optimality condition of the optimization problem (P1.1) is satisfied if the following equation holds:
\begin{eqnarray}
\begin{aligned}
\mathbf \Omega_{k,i}^{-1}\left(\{\bm g_{n,j}^s\}_{(n,j)\in\mathcal Q}\right)&\mathbf \Psi_{k,i}\left(\{\bm g_{n,j}^s\}_{(n,j)\in\mathcal Q}\right)\bm g_{k,i}^s\\&=\lambda \left(\{\bm g_{n,j}^s\}_{(n,j)\in\mathcal Q}\right)\bm g_{k,i}^s,
\label{eq:l1}
\end{aligned}
\end{eqnarray}
where $\mathbf \Psi_{k,i}\left(\cdot\right), \mathbf \Omega_{k,i}\left(\cdot\right)$, and $\lambda (\cdot)$ are defined in (\ref{eq:phi}).
\begin{figure*}[!t]
	\vspace*{-8pt}
	\normalsize
	\begin{eqnarray}
	\begin{aligned}
	&\mathbf \Psi_{k,i}\left(\{\bm g_{n,j}^s\}_{(n,j)\in\mathcal Q}\right)=\frac{\exp\left( \sum\limits_{j=1}^2\log_2\left(\frac{(\bm g_{k,j}^s)^H\mathbf {\Gamma}_{k,j}^u\bm g_{k,j}^s}{(\bm g_{k,j}^s)^H\mathbf {\Gamma}_{k,j}^d\bm g_{k,j}}^s\right)^{-\frac{1}{\alpha}}\right)}{\sum\limits_{n\in\mathcal K}\exp\left( \sum\limits_{j=1}^2\log_2\left(\frac{(\bm g_{n,j}^s)^H\mathbf {\Gamma}_{n,j}^u\bm g_{n,j}^s}{(\bm g_{n,j}^s)^H\mathbf {\Gamma}_{n,j}^d\bm g_{n,j}^s}\right)^{-\frac{1}{\alpha}}\right)}\times \frac{{\mathbf {\Gamma}_{k,i}^u}}{(\bm g_{k,i}^s)^H\mathbf {\Gamma}_{k,i}^u\bm g_{k,i}^s}\times\lambda_1(\{\bm g_{n,j}^s\}_{(n,j)\in\mathcal Q}),
	\\&\mathbf \Omega_{k,i}\left(\{\bm g_{n,j}^s\}_{(n,j)\in\mathcal Q}\right) =\frac{\exp\left( \sum\limits_{j=1}^2\log_2\left(\frac{(\bm g_{k,j}^s)^H\mathbf {\Gamma}_{k,j}^u\bm g_{k,j}^s}{(\bm g_{k,j}^s)^H\mathbf {\Gamma}_{k,j}^d\bm g_{k,j}^s}\right)^{-\frac{1}{\alpha}}\right)}{\sum\limits_{n\in\mathcal K}\exp\left( \sum\limits_{j=1}^2\log_2\left(\frac{(\bm g_{n,j}^s)^H\mathbf {\Gamma}_{n,j}^u\bm g_{n,j}^s}{(\bm g_{n,j}^s)^H\mathbf {\Gamma}_{n,j}^d\bm g_{n,j}^s}\right)^{-\frac{1}{\alpha}}\right)} \times \frac{{\mathbf {\Gamma}_{k,i}^d}}{(\bm g_{k,i}^s)^H\mathbf {\Gamma}_{k,i}^d\bm g_{k,i}^s}\times\lambda_2(\{\bm g_{n,j}^s\}_{(n,j)\in\mathcal Q}),
	\\&\lambda (\{\bm g_{n,j}^s\}_{(n,j)\in\mathcal Q}) =\frac{\lambda_1(\{\bm g_{n,j}^s\}_{(n,j)\in\mathcal Q})}{\lambda_2(\{\bm g_{n,j}^s\}_{(n,j)\in\mathcal Q})}= \log\left(\frac{1}{K}\sum\limits_{n\in\mathcal K}\exp\left( \sum\limits_{j=1}^2\log_2\left(\frac{(\bm g_{n,j}^s)^H\mathbf {\Gamma}_{n,j}^u\bm g_{n,j}^s}{(\bm g_{n,j}^s)^H\mathbf {\Gamma}_{n,j}^d\bm g_{n,j}^s}\right)^{-\frac{1}{\alpha}}\right)\right)^{-\alpha}.
	\label{eq:phi}
	\end{aligned}
	\end{eqnarray}
	\hrulefill
	\vspace*{-8pt}
\end{figure*}

{\it Proof:} Please refer to Appendix \ref{appendix:l1}.

It is observed that if $\bm g_{k, i}^s$ satisfies the condition (\ref{eq:l1}), $\bm g_{k, i}^s$ is a stationary point of (P1.1) where the gradient is zero. According to \cite{RN461,RN460}, (\ref{eq:l1}) is a form of the eigenvector problem for matrix $\mathbf \Omega_{k, i}^{-1}\left(\{\bm g_{n,j}^s\}_{(n,j)\in\mathcal Q}\right)\mathbf \Psi_{k, i}\left(\{\bm g_{n,j}^s\}_{(n,j)\in\mathcal Q}\right)$. This problem can be formulated as an eigenvector-dependent nonlinear eigenvalue problem (NEPv) which is a generalized version of an eigenvalue problem. In NEPv, a matrix can vary based on an eigenvector in a nonlinear fashion. In this context, $\mathbf \Omega_{k,i}^{-1}\left(\{\bm g_{n,j}^s\}_{(n,j)\in\mathcal Q}\right)\mathbf \Psi_{k,i}\left(\{\bm g_{n,j}^s\}_{(n,j)\in\mathcal Q}\right)$ is a nonlinear function of the eigenvector $\bm g_{k,i}^s$.
Note that $\lambda \left(\{\bm g_{n,j}^s\}_{(n,j)\in\mathcal Q}\right)$ is equivalent to the objective function (P1.1). Since there exist multiple $\bm g_{k, i}^s$ satisfying (\ref{eq:l1}), we can find the local optimal point that maximizes the objective function (P1.1) by finding the dominant eigenvector of the NEPv (\ref{eq:l1}), which maximizes $\lambda \left(\{\bm g_{n,j}^s\}_{(n,j)\in\mathcal Q}\right)$. It is worth mentioning that the power of $\|\bm g_{k, i}\|_2^2$ does not hurt the optimality since the receive beamforming power does not affect the spectral efficiency according to (\ref{eq:l1}). Therefore, without loss of generality, we can assume that $\|\bm g_{k, i}\|_2^2 = 1$ and (C1) can vanish from (P1.1). This leads to the following theorem.

\textbf{Theorem 1:} Let $\{\bm g_{k,i}^*\}_{(k,i)\in\mathcal Q}$ denote the local optimal solution of (P1.1). Then $\bm g_{k,i}^*$ is the eigenvector of $\mathbf \Omega_{k,i}^{-1}\left(\{\bm g_{n,j}^*\}_{(n,j)\in\mathcal Q}\right)\mathbf \Psi_{k,i}\left(\{\bm g_{n,j}^*\}_{(n,j)\in\mathcal Q}\right)$ satisfying
\begin{eqnarray}
\begin{aligned}
\mathbf \Omega_{k,i}^{-1}\left(\{\bm g_{n,j}^*\}_{(n,j)\in\mathcal Q}\right)&\mathbf \Psi_{k,i}\left(\{\bm g_{n,j}^*\}_{(n,j)\in\mathcal Q}\right)\bm g_{k,i}^*\\&=\lambda^*\bm g_{k,i}^* ,\forall  (k,i)\in\mathcal Q, 
\label{eq:t1}
\end{aligned}
\end{eqnarray}
where $\lambda^*=\lambda\left(\{\bm g_{n,j}^*\}_{(n,j)\in\mathcal Q}\right)$ is the corresponding eigenvalue.

However, it is difficult to find $\{\bm g_{k, i}^*\}_{(k, i)\in\mathcal Q}$  straightforward due to the nonlinear feature of the problem. According to \cite{RN460}, the self-consistent field (SCF) iteration is a natural and widely used approach for solving NEPv problems. Therefore, we propose Algorithm 1 to obtain $\{\bm g_{k,i}^*\}_{(k,i)\in\mathcal Q}$ in a computationally efficient fashion.

However, with the increasing number of devices, the computational complexity for the partial eigenvalue decomposition in Step 4 of Algorithm 1 becomes prohibitively high. To reduce complexity further, we introduce a power method to SCF iteration.
\begin{table}[ht]
	\normalsize
	\centering  
	\setlength{\tabcolsep}{1mm}{
		\begin{tabular}{l}  
			\hline  
			\textbf{Algorithm 1:} SCF iteration \\
			\hline	
			1: Initialize: $t=1$, $\bm g_{k,i}(0),\forall (k,i)\in\mathcal Q$ with unit norm,
			\\
			2: {Repeat} \\
			3: Calculate $\mathbf A_{k,i}(t)=\mathbf \Omega_{k,i}^{-1}(\{\bm g_{k,i}(t-1)\}_{(k,i)\in\mathcal Q})$\\\qquad\qquad\qquad\qquad\quad$\times\mathbf \Psi_{k,i}(\{\bm g_{k,i}(t-1)\}_{(k,i)\in\mathcal Q})$,\\
			4: Compute the partial eigenvalue decomposition\\ \quad $\mathbf A_{k,i}(t)\bm g_{k,i}(t)=\lambda(t)\bm g_{k,i}(t)$,
			\\
			5: $t\gets t+1$,\\
			6: Until $\sum\limits_{(k,i)\in\mathcal Q}\|\bm g_{k,i}(t)-\bm g_{k,i}(t-1)\|_2\le \kappa_1$,\\
			7: {Return} $\lambda(t),\bm g_{k,i}(t),\forall (k,i)\in\mathcal Q$.\\
			\hline
		\end{tabular}
	}
\end{table}

\textbf{Lemma 2:} For a Hermitian matrix $\mathbf A$, there exists a nonzero vector $\bm u_0$ such that the sequence of vectors given by
\begin{eqnarray}
\begin{aligned}
\{\bm u_0,\mathbf A\bm u_0, \mathbf A^2\bm u_0,\ldots,\mathbf A^{t'}\bm u_0,\ldots\}
\end{aligned}
\end{eqnarray}
approaches a multiple of the dominant eigenvector of $\mathbf A$.

{\it Proof:} Please refer to Appendix \ref{appendix:l2}.

Since we assume that $\|\bm g_{k, i}\|_2^2 = 1$, according to Lemma 2, we can replace Step 4 in Algorithm 1 by `$\bm g_{k,i}(t)=\frac{\mathbf A_{k,i}^{t'}(t)\bm u_0}{\|\mathbf A_{k,i}^{t'}(t)\bm u_0\|_2}$'. At this point, the algorithm requires two layers of iterations, i.e., the outer iteration represented by index $t$ and the inner iteration represented by index $t'$. To avoid this issue, we further give the following theorem.

\textbf{Theorem 2:} Given a nonzero unit vector $\bm g_{k,i}^s(t-1)$ and $t'> 0$, 
there is
\begin{eqnarray}
\begin{aligned}
\bm g_{k,i}^s(t+t'-1)&=\frac{\mathbf A_{k,i}(t+t'-1)\bm g_{k,i}^s(t+t'-2)}{\|\mathbf A_{k,i}(t+t'-1)\bm g_{k,i}^s(t+t'-2)\|_2}\\&=\frac{\mathbf A_{k,i}^{t'}(t)\bm g_{k,i}^s(t-1)}{\|\mathbf A_{k,i}^{t'}(t)\bm g_{k,i}^s(t-1)\|_2},
\label{eq:t2}
\end{aligned}
\end{eqnarray} 
where $\mathbf A_{k,i}(t)=\mathbf \Omega_{k,i}^{-1}(\{\bm g_{k,i}^s(t-1)\}_{(k,i)\in\mathcal Q})\mathbf \Psi_{k,i}(\{\bm g_{k,i}^s(t-1)\}_{(k,i)\in\mathcal Q})$.

{\it Proof:} Please refer to Appendix \ref{appendix:t3}.

According to Theorem 2, we can incorporate the inner iteration into the outer one by modifying Step 4 as follows:  `$\bm g_{k,i}(t)=\frac{\mathbf A_{k,i}(t)\bm g_{k,i}(t-1)}{\|\mathbf A_{k,i}(t)\bm g_{k,i}(t-1)\|_2}$'.
With this adjustment, we can present the following GPI algorithm.
\begin{table}[ht]
	\normalsize
	\centering  
	\setlength{\tabcolsep}{1mm}{
		\begin{tabular}{l}  
			\hline  
			\textbf{Algorithm 2:} The generalized power iteration (GPI)\\
			\hline	
			1: Initialize: $t=1$, $\bm g_{k,i}(0),\forall (k,i)\in\mathcal Q$ with unit norm.
			\\
			2: {Repeat} \\
			3: Calculate $\mathbf A_{k,i}(t)=\mathbf \Omega_{k,i}^{-1}(\{\bm g_{k,i}(t-1)\}_{(k,i)\in\mathcal Q})$\\\qquad\qquad\qquad\qquad\quad$\times\mathbf \Psi_{k,i}(\{\bm g_{k,i}(t-1)\}_{(k,i)\in\mathcal Q})$,\\
			4: Compute the partial eigenvalue decomposition \\ \quad$\bm g_{k,i}(t)=\frac{\mathbf A_{k,i}(t)\bm g_{k,i}(t-1)}{\|\mathbf A_{k,i}(t)\bm g_{k,i}(t-1)\|_2}$,
			\\5: Compute $\lambda(t)$ according to (\ref{eq:phi}),
			\\
			6: $t\gets t+1$,\\
			7: Until $\frac{\sum\limits_{(k,i)\in\mathcal Q}\|\bm g_{k,i}(t)-\bm g_{k,i}(t-1)\|_2}{\sum\limits_{(k,i)\in\mathcal Q}\|\bm g_{k,i}(t-1)\|_2}\le \kappa_1$, \\
			8: {Return} $\lambda(t),\bm g_{k,i}(t),\forall (k,i)\in\mathcal Q$.\\
			\hline
		\end{tabular}
	}
	\vspace{-0.6cm}
\end{table}

\subsection{Optimizing Phase-shift Beamforming}
\label{opt_V}
In this part, our objective is to optimize phase-shift beamforming given receive beamforming, power allocation, and decoding order. However, due to the constraint (C2), the phase-shift beamforming optimization cannot adopt the GPI method. Therefore, we resort to employing SDP to solve this problem.  Specifically, when $(k,i)\in\mathcal Q_l$, we define 
\begin{align}
&u_{k,i}(\bm g_{k,i},\mathbf V,\bm p)\nonumber \triangleq\\&\log_2\left(\sum\limits_{(n,j)\in\underline{\mathcal Q}_l}p_{n,j}\text{tr}(\bm g_{k,i}^H\mathbf H_n\mathbf V\mathbf H_n^H\bm g_{k,i} )+\text{tr}(\bm g_{k,i}^H\bm g_{k,i})\sigma^2\right),\\
&d_{k,i}(\bm g_{k,i},\mathbf V,\bm p)\nonumber
\triangleq\\&\log_2\left(\sum\limits_{\substack{(n,j)\in\underline{\mathcal Q}_l,\\(n,j)\neq(k,i)}}p_{n,j}\text{tr}(\bm g_{k,i}^H\mathbf H_n\mathbf V\mathbf H_n^H\bm g_{k,i} )+\text{tr}(\bm g_{k,i}^H\bm g_{k,i})\sigma^2\right).
\end{align} 
where $\bm p=[p_{1,1},p_{1,2},\ldots,p_{K,1},p_{K,2}]$. Given $\mathcal Q_l, \bm p$, and $\mathbf g_{k,i},\forall (k,i)\in\mathcal Q$, we introduce an auxiliary variable $r$ and covert (P0) into the following form,
\begin{eqnarray}
\begin{aligned}
\text{(P2)}\quad \max_{\mathbf V, r}& \quad r\\
\text{s.t.} \quad &\text{(C2)},
\text{(C3)},
\\& \text{(C6)}\sum\limits_{i=1}^2 u_{k,i}-d_{k,i}\ge r, \forall k\in\mathcal K.
\end{aligned}\label{eq:p21}
\end{eqnarray}
Note that the functions $u_{k,i}$ and $d_{k,i}$ are concave w.r.t. $\mathbf V$. However, the concavity of $d_{k, i}$ makes the optimization problem no-convex. To tackle this issue, we employ the iterative successive convex approximation (SCA) method. Specifically, we use SCA to linearly approximate $d_{k,i}$ as follows
\begin{eqnarray}
\begin{aligned}	
d_{k,i}(\mathbf V)&\leq d_{k,i}(\mathbf V^t)+\text{tr}\left(\left(\nabla_{\mathbf V}d_{k,i}(\mathbf V^t)\right)^T\left(\mathbf V-\mathbf V^t\right)\right)\\&\triangleq d_{k,i}^t(\mathbf V),\label{eq:appr1}
\end{aligned}
\end{eqnarray}
where  $\nabla_{\mathbf V}d_{k,i}(\mathbf V^t)={\left(\sum\limits_{\substack{(n,j)\in\underline{\mathcal Q}_l,\\(n,j)\neq(k,i)}}p_{n,j}\mathbf H_n^H\bm g_{k,i}\bm g_{k,i}^H\mathbf H_n\right)^T}\biggl/$\\${\left(\sum\limits_{\substack{(n,j)\in\underline{\mathcal Q}_l,\\(n,j)\neq(k,i)}}p_{n,j}\text{tr}(\mathbf H_n\mathbf V^t\mathbf H_n^H\bm g_{k,i}\bm g_{k,i}^H )+\text{tr}(\bm g_{k,i}\bm g_{k,i}^H)\sigma^2\right)\ln2}$. $\mathbf V^t$ is the local feasible point in the $t$th iteration. Eq.(\ref{eq:appr1}) gives an upper bound of $d_{k,i}$ by its first-order Taylor expansion. Therefore, the optimization problem can be approximately transformed into
\begin{eqnarray}
\begin{aligned}
\text{(P2.1)}\quad \max_{\mathbf V, r}& \quad r\\
\text{s.t.} \quad &\text{(C2)},
\text{(C3)},
\\&(\overline{\text{C6}})\quad\sum\limits_{i=1}^2 u_{k,i}-d_{k,i}^t(\mathbf V)\ge r, \forall k\in\mathcal K.
\end{aligned}\label{eq:p1.2}
\end{eqnarray}
However, due to the non-convex rank constraint in (C3), problem (P2.1) is still a non-convex problem. To address this problem, we exploit the penalty-based method \cite{RN416} to handle the rank constraint. To be specific, 
\begin{align}
\text{rank}(\mathbf V)= 1\Rightarrow \text{tr}(\mathbf V)-\|\mathbf V\|_2=0.
\end{align} 
Then, we incorporate the constraint $\text{tr}(\mathbf V)-\|\mathbf V\|_2=0$ into the objective function (P2.1) by introducing a positive penalty parameter $\rho_1$, and obtain the problem (P2.2) as follows,
\begin{eqnarray}
\begin{aligned}
\text{(P2.2)}\quad\max_{\mathbf V,r}& \quad  r -\frac{1}{2\rho_1}(\text{tr}\left(\mathbf V\right)-\|\mathbf V\|_2)\\
\text{s.t.}&\quad\text{(C2)},(\overline{\text{C6}}),
\\&\quad(\overline{\text{C3}}) \quad\mathbf V\succeq\mathbf 0.
\label{eq:p1.3}
\end{aligned}
\end{eqnarray}
According to Theorem 3, problem (P2.2) can obtain a rank-one solution when $\rho_1$ is sufficiently small. 

\textbf{Theorem 3:} Let $\mathbf V^s$ denote the optimal solution of (P2.2) with penalty parameter $\rho_s$. When $\rho_s$ is sufficiently small, i.e., $\rho_s\to 0$, then any limit point $\bar{\mathbf V}$ of the sequence $\{\mathbf V^s\}$ is an optimal solution of problem (P2.1).

{\it{Proof:}} Please refer to Appendix \ref{appendix:t1}.

Note that the convexity of $\|\mathbf V\|_2$ makes problem (P2.2) still non-convex. Therefore, we replace $\|\mathbf V\|_2$ with a lower bound given by a first-order Taylor expansion of $\|\mathbf V\|_2$, i.e.,
\begin{align}
\|\mathbf V\|_2\geq \|\mathbf V^t\|_2+\text{tr}\left(\bm\lambda_{\text{max}}^t\bm(\bm\lambda_{\text{max}}^t)^H\left(\mathbf V-\mathbf V^t\right)\right),
\end{align}
where $\bm\lambda_{\text{max}}^t$ represents the  eigenvector corresponding to the largest eigenvalue of $\mathbf V^t$. Then we can approximate problem (P2.2) as
\begin{eqnarray}
\begin{aligned}
\text{(P2.3)}\quad\max_{\mathbf V,r}& \quad  r -\frac{1}{2\rho_1}\left(\text{tr}\left(\mathbf V\right)-\|\mathbf V^t\|_2\right.\\&\quad\left. -\text{tr}\left(\bm\lambda_{\text{max}}^t\bm(\bm\lambda_{\text{max}}^t)^H\left(\mathbf V-\mathbf V^t\right)\right)\right)\\
\text{s.t.} &\quad\text{(C2)},(\overline{\text{C3}}),(\overline{\text{C6}}).
\end{aligned}
\end{eqnarray}
It is observed that (P2.3) is an SDP problem, which can be efficiently solved by off-the-shelf solvers such as CVX.

\subsection{Optimizing Power Allocation }
\label{se:opa}
Given $\bm g_{k,i}$, $\mathbf V$, and $\mathcal Q_l$, (P0) can be written as  
\begin{eqnarray}
\begin{aligned}
\text{(P3)}\quad \max_{p_{k,i},r}& \quad r\\
\text{s.t.}&\quad \text{(C4)},\text{(C5)},\text{(C6)}.
\end{aligned}\label{eq:p1.81}
\end{eqnarray}
We can approximate $d_{k,i}(\bm p)$ by Taylor expansion and obtain 
\begin{eqnarray}
\begin{aligned}	
d_{k,i}(\bm p)&\leq d_{k,i}(\bm p^t )+\text{tr}\left(\left(\nabla_{\bm p}d_{k,i}\left(\bm p^t\right)\right)^T\left(\bm p-\bm p^t\right)\right)\\&\triangleq d_{k,i}^t(\bm p),
\end{aligned}\label{eq:appr12}
\end{eqnarray}
with
\begin{eqnarray}
\begin{aligned}
&\nabla_{\bm p}d_{k,i}\left(\bm p^t\right)=\\& \frac{\mathfrak{T}_{k,i}}{\left(\sum\limits_{\substack{(n,j)\in\underline{\mathcal Q}_l,\\(n,j)\neq(k,i)}}p_{n,j}^t\text{tr}(\bm g_{k,i}^H\mathbf H_n\mathbf V\mathbf H_n^H\bm g_{k,i})+\text{tr}(\bm g_{k,i}^H\bm g_{k,i})\sigma^2\right)\ln2},\nonumber
\end{aligned}
\end{eqnarray}
where
\begin{eqnarray}
\begin{aligned}
&\mathfrak{T}_{k,i}=\left[\overbrace{\text{tr}\left(\bm g_{k,i}^H\mathbf H_1\mathbf V\mathbf H_1^H\bm g_{k,i}\right),\text{tr}\left(\bm g_{k,i}^H\mathbf H_1\mathbf V\mathbf H_1^H\bm g_{k,i}\right),\ldots,}^{2(k-1)\quad \text{elements}}\right.\\&\left.|i-1|*\text{tr}\left(\bm g_{k,i}^H\mathbf H_k\mathbf V\mathbf H_k^H\bm g_{k,i}\right),|i-2|*\text{tr}\left(\bm g_{k,i}^H\mathbf H_k\mathbf V\mathbf H_k^H\bm g_{k,i}\right),\right.\\&\qquad\left.\underbrace{\ldots,\text{tr}\left(\bm g_{k,i}^H\mathbf H_K\mathbf V\mathbf H_K^H\bm g_{k,i}\right),\text{tr}\left(\bm g_{k,i}^H\mathbf H_K\mathbf V\mathbf H_K^H\bm g_{k,i}\right)}_{2(K-k)\quad \text{elements}}\right]^T.\nonumber
\end{aligned}
\end{eqnarray}
$\bm p^t$ denotes the local feasible point in the $t$th iteration. Then we can convert (P3) into
\begin{eqnarray}
\begin{aligned}
\text{(P3.1)}\quad \max_{p_{k,i},r}& \quad r\\
\text{s.t.}&\quad \text{(C4)},\text{(C5)},
\\&\quad(\overline{\text{C6}})\quad \sum\limits_{i=1}^2 u_{k,i}-d_{k,i}^t(\bm p)\ge r, \forall k\in\mathcal K.
\end{aligned}\label{eq:p1.9}
\end{eqnarray}
(P3.1) is a linear programming (LP) problem and it can be solved by existing CVX.

\subsection{Optimizing Decoding Order}
\label{se:odo}
The group decoding order is significant for SGD because it decides the minimum rate of the system. Therefore, the objective of this section is to identify the group decoding order for different devices under the max-min constraints. Given receive beamforming $\bm g_{k,i}$, phase-shift beamforming $\mathbf V$ and power allocation $\bm p$, (P0) can be written as 
\begin{eqnarray}
\begin{aligned}
\text{(P4)}\quad \max_{\mathcal Q_l}& \quad \min_{k\in\mathcal K} \sum_{i=1}^2 r_{k,i}.
\end{aligned}\label{eq:p1.8}
\end{eqnarray} 
It is observed that (P4) is a combinatorial optimization problem, which is challenging to solve using exact methods. In this section, we propose a coordinated greedy grouping scheme based on \cite{RN439}, which improves the minimum rate by moving corresponding devices from their group to the next one in order.

Based on Section \ref{SGD}, in the $l$th stage of the SGD, sub-massages in group $\mathcal Q_l$ are decoded. Since the interference experienced by sub-messages in the $(l+1)$-th group is typically smaller than that in the $l$-th group, the sub-message can enhance its rate by moving it to the group that is decoded later in the process. Moreover, since the CSIT is the same for sub-messages of the same device,  we distribute the sub-messages of the same device across different groups to reduce the influence of CSIT. The greedy grouping algorithm is summarized in Algorithm 3.
We can outline the procedure as follows:
\begin{enumerate}
	\item At initialization, set all the devices in $\mathcal Q_1$ and leave the rest of the groups empty (step 1). 
	\item We compute the rate of each sub-message by (\ref{eq:sr}) and find the device $k^*$ with the minimum rate (step 3). Then move the sub-message $(k^*,i^*)$ with the lower rate of device $k^*$ to the next group (steps 6-8).
	\item To prevent sub-messages of the same device from being in the same group, we adopt the following method. Assume $(k^*,i^*)\in \mathcal Q_l$, if the other sub-message $(k^*,3-i^*)$ of $k^*$ is in the $(l+1)$-th group, move $(k^*,i^*)$ to the $(l+2)$-th group (steps 9-11).
	\item The iteration will stop under the following conditions: i) the iteration count reaches a sufficient number (step 16); ii) $(k^*,i^*)\in \mathcal Q_L$, since the $L$-group is the last group to be decoded, the minimum rate cannot be improved further (step 14); iii) $(k^*,i^*)\in \mathcal Q_{L-2}$ and $(k^*,3-i^*)\in \mathcal Q_L$. In this case, to avoid interference from the same device, further improvement is not possible (step 12).
\end{enumerate}

\begin{table}[ht]
	\normalsize
	\centering  
	\setlength{\tabcolsep}{1mm}{
		\begin{tabular}{l}  
			\hline  
			\textbf{Algorithm 3:} Greedy Grouping Algorithm \\
			\hline	
			1: Initialize: $t=0$, $\mathcal Q_1^t=\mathcal Q$, $\mathcal Q_2^t=\ldots=\mathcal Q_L^t=\emptyset$,\\ \quad$T_{\text{max}}=KI(L-1)$, $T_{\text{stop}}=0$.
			\\
			2: {Repeat} \\
			3: Calculate\quad$k^*=\arg\min\limits_{k\in\mathcal K}\sum\limits_{i=1}^2 r_{k,i},$\\ \qquad\quad\qquad$(k^*,i^*)=\arg\min\limits_{i=1,2} r_{k^*,i}$,\\
			4: Assume $(k^*,i^*)\in\mathcal Q_l^t$, \\
			5:  if $l<L$,\\
			6: \quad if $(k^*,3-i^*)\notin\mathcal Q_{l+1}^t$,\\
			7: \qquad $\mathcal Q_l^{t+1}\gets \mathcal Q_l^t\backslash (k^*,i^*)$, $\mathcal Q_{l+1}^{t+1}\gets \mathcal Q_{l+1}^t\cup (k^*,i^*)$,\\
			8:\qquad $\mathcal Q_{l'}^{t+1}\gets\mathcal Q_{l'}^t, (1\le l'\le L, l'\ne l,l+1)$,\quad $t\gets t+1$,\\
			9: \quad else if $l<L-1$, \\
			10: \qquad $\mathcal Q_l^{t+1}\gets \mathcal Q_l^t\backslash (k^*,i^*)$, $\mathcal Q_{l+2}^{t+1}\gets \mathcal Q_{l+2}^t\cup (k^*,i^*)$,\\
			11:\qquad $\mathcal Q_{l'}^{t+1}\gets\mathcal Q_{l'}^t, (1\le l'\le L, l'\ne l,l+2)$,\quad $t\gets t+1$,\\
			12: \quad else $T_{\text{stop}}=1$,\\
			13:\quad end\\
			14:   else $T_{\text{stop}}=1$,\\
			15: end\\
			16: Until  $t>T_{\text{max}}$ or $T_{\text{stop}}=1$,\\
			17: $\mathcal Q_1\gets\mathcal Q_1^{t-1},\ldots,\mathcal Q_L\gets \mathcal Q_L^{t-1}$,\\
			18: {Return} $\mathcal Q_1,\ldots,\mathcal Q_L$.\\
			\hline
		\end{tabular}
	}
\end{table}

\subsection{The Overall Fairness Optimization Algorithm in The IRS-aided RSMA System}

The overall AO algorithm proposed in this section is summarized in Algorithm 4.  Note that Algorithm 2 can obtain the maximum value
of (P1.1). (P2.3) and (P3.1) serve as lower bounds for the optimal values of (P2.2) and (P3), respectively. By iteratively solving (P1.1), (P2.3), and (P3.1), we can progressively tighten these lower bounds. Algorithm 3 is a greedy algorithm that guarantees a non-decreasing minimum rate. In this way, the objective
values achieved by the sequence $\{\bm g_{k,i}(t), \mathbf V(t), p_{k,i}(t), \mathcal Q_l^{t}\}_{t\in\mathbb N}$ form a non-decreasing sequence that converges to a stationary value.

\begin{table}[ht]
	\normalsize
	\centering  
	\setlength{\tabcolsep}{1mm}{
		\begin{tabular}{l}  
			\hline  
			\textbf{Algorithm 4:} The AO Algorithm\\
			\hline	
			1: Initialize: $t=1$, $\mathcal Q_1^0=\mathcal Q$, $\mathcal Q_2^0=\ldots=\mathcal Q_L^0=\emptyset$, randomly\\ \quad construct $\bm g_{k,i}(0)$ with unit norm, and  $p_{k,i}(0)$  with power\\ \quad 
			constraint, $\mathbf V(0)$ with phase constraint.
			\\
			2: {Repeat} \\
			3: Find $\bm g_{k,i}(t)$ by Algorithm 2 with the given $\mathbf V(t-1)$,\\ \quad $p_{k,i}(t-1)$, and $\mathcal Q_l^{t-1}$,\\
			4: Update $\mathbf V(t)$ by solving (P2.3) with the given $\bm g_{k,i}(t)$, \\ \quad $p_{k,i}(t-1)$, and $\mathcal Q_l^{t-1}$,
			\\5: Update $p_{k,i}(t)$ by solving (P3.1) with the given $\bm g_{k,i}(t)$,\\ \quad  $\mathbf V(t)$, and $\mathcal Q_l^{t-1}$,
			\\
			6: Update $\mathcal Q_l^{t}$ by Algorithm 3 with the given $\bm g_{k,i}(t)$, \\ \quad $\mathbf V(t)$, and $p_{k,i}(t)$,\\
			7: Compute $r_{\text{min}}(t)=\sum\limits_{i=1}^2 r_{k,i}(t),\forall k\in \mathcal K$,\\
			8: $t\gets t+1$,\\
			9: Until $\frac{r_{\text{min}}(t)-r_{\text{min}}(t-1)}{r_{\text{min}}(t-1)}\le \kappa_2$, \\
			10: {Return} $\bm g_{k,i}(t)$, $\mathbf V(t)$, $p_{k,i}(t)$, and $\mathcal Q_l^{t}$.\\
			\hline
		\end{tabular}
	}
	\vspace{-0.6cm}
\end{table}

\subsection{Computational Complexity Analysis}
\label{Com}
The complexity of the proposed scheme mainly depends on iteratively solving (P1.1) by Algorithm 2, (P2.3) by SDP, (P3.3) by LP, and (P4) by Algorithm 3. The total computational
complexity of Algorithm 2 is dominated by the calculation of $\mathbf \Omega_{k,i}^{-1}\left(\{\bm g_{n,j}\}_{(n,j)\in\mathcal Q}\right)$. Since  $\mathbf \Omega_{k,i}\left(\{\bm g_{n,j}\}_{(n,j)\in\mathcal Q}\right)$ is a $M\times M$ matrix, the inverse operation requires a complexity of $\mathcal O(M^3)$. The complexity of solving (P1.1) is $\mathcal O(KIM^3)$.
In each iteration, using the interior point method, the computational complexity of solving (P2.3) and (P3.3) is $\mathcal{O}(KN^{3.5}+N^{2.5}K^2+\sqrt{N}K^3)$ and $\mathcal O(KI)$, respectively. (P4) is solved by Algorithm 3, which is a kind of greedy algorithm, so its complexity can be represented by $\mathcal O(L)$.
Therefore, the computational complexity of the proposed scheme is $\mathcal O(t(KN^{3.5}+N^{2.5}K^2+\sqrt{N}K^3+KIM^{3}+KI+L))$, where $t$ is the number of iterations.

\section{Resource Allocation for Fairness with Imperfect CSIT}
\label{imCSI}
In the above, we discussed the fair rate adaption when CSIT is known. However, in actual communication systems, the BS needs to estimate CSIT, which is often inaccurate. Therefore, it is significant to study robust uplink RSMA communication. In this section, we focus on fair rate adaptation for the IRS-aided uplink network with imperfect CSIT. Considering the statistical information of channels, we derive a lower bound on the minimum rate and maximize the lower bound.

\subsection{Problem Formulation}

Accordingly, under the premise that the BS has imperfect CSIT, i.e., $\mathbf H_k = \hat{\mathbf H}_k + \mathbf E_k, \forall k\in \mathcal K$, the ergodic spectral efficiency of the sub-message is obtained as (\ref{eq:Er}),
\begin{figure*}[!t]
	\vspace*{-8pt}
	\normalsize
	\begin{eqnarray}	
	\begin{aligned}
	\bar r_{k,i}
	=\mathbb E_{\hat{\mathbf H}_k}\left[\left.\mathbb E_{\mathbf E_k}\left[\log_2\left(1+\frac{p_{k,i}\|\bm g_{k,i}^H\mathbf H_k\bm v\|_2^2}{\sum\limits_{\substack{(n,j)\in\underline{\mathcal Q}_l,\\(n,j)\neq(k,i)}}p_{n,j}\|\bm g_{k,i}^H\mathbf H_n\bm v\|_2^2+\|\bm g_{k,i}\|_2^2\sigma^2}\right)\right|\hat{\mathbf H}_k\right]\right].
	\label{eq:Er}
	\end{aligned}
	\end{eqnarray}
	\vspace*{-8pt}
\end{figure*}
where $\mathbf E_k$ denotes the CSIT estimation error. Our main goal is to optimize the minimum rate with imperfect CSIT in each fading block. Without loss of generality, assume that $\hat{\mathbf H}_k, \forall k\in \mathcal K$ is given. The instantaneous rate of $(k,i)$ is defined as
\begin{eqnarray}	
\begin{aligned}
&\bar r_{k,i}^{\text{ins}}=\\&\left.\mathbb E_{\mathbf E_k}\left[\log_2\left(\frac{\sum\limits_{(n,j)\in\underline{\mathcal Q}_l}p_{n,j}\|\bm g_{k,i}^H\mathbf H_n\bm v\|_2^2+\|\bm g_{k,i}\|_2^2\sigma^2}{\sum\limits_{\substack{(n,j)\in\underline{\mathcal Q}_l,\\(n,j)\neq(k,i)}}p_{n,j}\|\bm g_{k,i}^H\mathbf H_n\bm v\|_2^2+\|\bm g_{k,i}\|_2^2\sigma^2}\right)\right|\hat{\mathbf H}_k\right].\nonumber
\end{aligned}
\end{eqnarray}
Considering multiple fading blocks, the instantaneous rate and the ergodic rate are connected by
$\bar r_{k,i}=\mathbb E_{\hat{\mathbf H}_k}\left[\bar r_{k,i}^{\text{ins}}\right]$.

To obtain the closed-form expectation considering CSIT error, we rewrite the received signal (\ref{eq:x}) with the CSIT error term as (\ref{eq:x2}),
\begin{figure*}[!t]
	\vspace*{-8pt}
	\normalsize
	\begin{eqnarray}	
	\begin{aligned}
	\hat{x}_{k,i}&=\bm g_{k,i}^H\hat{\mathbf H}_k\bm v \sqrt{p_{k,i}} x_{k,i}+\bm g_{k,i}^H\sum\limits_{\substack{(n,j)\in\underline{\mathcal Q}_l,\\(n,j)\neq (k,i)}}\hat{\mathbf H}_n\bm v\sqrt{p_{n,j}} x_{n,j} + \bm g_{k,i}^H\mathbf E_k\bm v \sqrt{p_{k,i}} x_{k,i}+\bm g_{k,i}^H\sum\limits_{\substack{(n,j)\in\underline{\mathcal Q}_l,\\(n,j)\neq (k,i)}}\mathbf E_n\bm v\sqrt{p_{n,j}} x_{n,j} + \bm g_{k,i}^H\bm w\\&=\bm g_{k,i}^H\tilde{\mathbf V}\hat{\bm h}_k \sqrt{p_{k,i}} x_{k,i}+\bm g_{k,i}^H\sum\limits_{\substack{(n,j)\in\underline{\mathcal Q}_l,\\(n,j)\neq (k,i)}}\tilde{\mathbf V}\hat{\bm h}_n\sqrt{p_{n,j}} x_{n,j} + \bm g_{k,i}^H\tilde{\mathbf V}\tilde{\bm e}_k \sqrt{p_{k,i}} x_{k,i}+\bm g_{k,i}^H\sum\limits_{\substack{(n,j)\in\underline{\mathcal Q}_l,\\(n,j)\neq (k,i)}}\tilde{\mathbf V}\tilde{\bm e}_n \sqrt{p_{n,j}} x_{n,j} + \bm g_{k,i}^H\bm w.
	\label{eq:x2}
	\end{aligned}
	\end{eqnarray} 
	\hrulefill
	\vspace*{-8pt}
\end{figure*}
where $\hat{\bm h}_k=\left[[\hat{\mathbf H}_k]_{1:},[\hat{\mathbf H}_k]_{2:},\ldots,[\hat{\mathbf H}_k]_{M:}\right]^T\in\mathbb C^{M(N+1)\times 1}$, $\tilde{\bm e}_k=\left[[\mathbf E_k]_{1:},[\mathbf E_k]_{2:},\ldots,[\mathbf E_k]_{M:}\right]^T\in\mathbb C^{M(N+1)\times 1}$, and $\tilde{\mathbf V} = [\tilde{\bm v}_1^T;\ldots;\tilde{\bm v}_M^T]\in \mathbb C^{M\times M(N+1) }$ with 
\begin{eqnarray}	
\begin{aligned}
\tilde{\bm v}_m = [\underbrace{\bm 0_{N+1};\ldots;\bm 0_{N+1}}_{m-1\,\text{ zero vectors}};\bm v;\underbrace{\bm 0_{N+1};\ldots;\bm 0_{N+1}}_{M-m\text{ zero vectors}}]\in \mathbb C^{M(N+1)\times 1 }.\nonumber
\end{aligned}
\end{eqnarray} 
According to \cite{RN451}, define
\begin{eqnarray}
\begin{aligned}
\mathbb E_{\bm e_k}\left[\tilde{\bm e}_k\tilde{\bm e}_k^H\right]=\mathbf{\Phi}_k = \mathbf{\Sigma}_k-\mathbf{\Sigma}_k\left(\mathbf{\Sigma}_k+\frac{\sigma^2}{LP_{\text{max}}}\mathbf I \right)^{-1} \mathbf{\Sigma}_k,
\end{aligned}
\end{eqnarray}
where $\mathbf{\Sigma}_k=\mathbb E_{\bm h_k}\left[\tilde{\bm h}_k\tilde{\bm h}_k^H\right] $ with $\tilde{\bm h}_k=\left[[\mathbf H_k]_{1:},[\mathbf H_k]_{2:},\ldots,[\mathbf H_k]_{M:}\right]^T$. $L$ is uplink training length. When $L, P_{\text{max}}\to \infty$, the error covariance $\mathbf{\Phi}_k\to 0$.
Treating the CSIT error in (\ref{eq:x2}) as independent Gaussian noise, a lower bound on the instantaneous rate (\ref{eq:rlower}) is derived,
\begin{figure*}[!t]
	\vspace*{-8pt}
	\normalsize
	\begin{align}	
	&\bar r_{k,i}^{\text{ins}}\ge
	\mathbb E_{\bm e_k}\left[\log_2\left(1+\frac{p_{k,i}\|\bm g_{k,i}^H\tilde{\mathbf V}\hat{\bm h}_k\|_2^2}{p_{k,i}\|\bm g_{k,i}^H\tilde{\mathbf V}\tilde{\bm e}_k\|_2^2+\sum\limits_{\substack{(n,j)\in\underline{\mathcal Q}_l,\\(n,j)\neq(k,i)}}p_{n,j}\left(\|\bm g_{k,i}^H\tilde{\mathbf V}\hat{\bm h}_n\|_2^2+\|\bm g_{k,i}^H\tilde{\mathbf V}\tilde{\bm e}_n\|_2^2\right)+\|\bm g_{k,i}\|_2^2\sigma^2}\right)\right]\nonumber
	\\&\overset{(a)}{\ge} 
	 \log_2\left(1+\frac{p_{k,i}\bm g_{k,i}^H\tilde{\mathbf V}\hat{\bm h}_k\hat{\bm h}_k^H\tilde{\mathbf V}^H\bm g_{k,i}}{p_{k,i}\bm g_{k,i}^H\tilde{\mathbf V}\mathbf{\Phi}_k\tilde{\mathbf V}^H\bm g_{k,i}+\sum\limits_{\substack{(n,j)\in\underline{\mathcal Q}_l,\\(n,j)\neq(k,i)}}p_{n,j}\left(\bm g_{k,i}^H\tilde{\mathbf V}\hat{\bm h}_n\hat{\bm h}_n^H\tilde{\mathbf V}^H\bm g_{k,i}+\bm g_{k,i}^H\tilde{\mathbf V}\mathbf{\Phi}_n\tilde{\mathbf V}^H\bm g_{k,i}\right)+\bm g_{k,i}^H\bm g_{k,i}\sigma^2}\right)\nonumber
	\\&=\bar r_{k,i}^{\text{lower}}.
	\label{eq:rlower}
	\end{align}
	\hrulefill
	\vspace*{-8pt}
\end{figure*}
where (a) follows Jensen’s inequality. 

To maximize the minimum rate among all devices, we formulate the max-min optimization problem as follows:
\begin{eqnarray}
\begin{aligned}
\text{(P5)} \max_{\mathcal Q_l,\bm v ,\bm g_{k,i},p_{k,i}}& \quad \min_{k\in\mathcal K} \sum\limits_{i=1}^2\bar r_{k,i}^{\text{lower}}\\
\text{s.t.}  &\quad\sum\limits_{i=1}^2 \|\bm g_{k,i}\|_2^2\leq P_{\text{max}}^{\text{b}}, \forall k\in \mathcal K, \\&\quad|[\bm v]_n| =1, n=1,\ldots,N,[\bm v]_{N+1}=1,\\&\quad \sum\limits_{i=1}^2 p_{k,i}\leq P_{\text{max}},\forall k\in\mathcal K,\\&\quad  p_{k,i}\ge 0,\forall (k,i)\in \mathcal Q\label{eq:p1co}.
\end{aligned}
\end{eqnarray}
As (P0) in Section \ref{P1}, (P5) is also an intractable non-convex problem, and we adopt the AO algorithm to solve this problem. In addition, since the power allocation and grouping order optimization are similar to Section \ref{se:odo} and \ref{se:opa}, respectively, we will not discuss them once again\footnote{The computational complexity of the scheme for the system with imperfect CSIT remains the same as that described in Section \ref{Com}.}.

\subsection{Optimizing Receive Beamforming}
\label{opt_G2}

Given decoding order, phase-shift beamforming, and power allocation, the subproblem of (P5) becomes 
\begin{eqnarray}
\begin{aligned}
\text{(P6)}\quad \max_{\bm g_{k,i}}& \quad \min_{k\in\mathcal K} \sum\limits_{i=1}^2\bar r_{k,i}^{\text{lower}}\\
\text{s.t.} & \quad \sum\limits_{i=1}^2 \|\bm g_{k,i}\|_2^2\leq P_{\text{max}}^{\text{b}}, \forall k\in\mathcal K.
\end{aligned}\label{eq:13b}
\end{eqnarray}
The lower bound $\bar r_{k,i}^{\text{lower}}$ can be rewritten as 
\begin{eqnarray}	
\begin{aligned}
\bar r_{k,i}^{\text{lower}}
= \log_2\left(\frac{\bm g_{k,i}^H\bar{\mathbf {\Gamma}}_{k,i}^u\bm g_{k,i}}{\bm g_{k,i}^H\bar{\mathbf {\Gamma}}_{k,i}^d\bm g_{k,i}}\right).\label{eq:rlow}
\end{aligned}
\end{eqnarray}
where 
\begin{eqnarray}	
\begin{aligned}
\bar{\mathbf {\Gamma}}_{k,i}^u&=\sum\limits_{(n,j)\in\underline{\mathcal Q}_l}p_{n,j}\left(\tilde{\mathbf V}\hat{\bm h}_n\hat{\bm h}_n^H\tilde{\mathbf V}^H+\tilde{\mathbf V}\mathbf{\Phi}_n\tilde{\mathbf V}^H\right)+\sigma^2\mathbf I,\\
\bar{\mathbf {\Gamma}}_{k,i}^d&=p_{k,i}\tilde{\mathbf V}\mathbf{\Phi}_k\tilde{\mathbf V}^H +\\& \sum\limits_{\substack{(n,j)\in\underline{\mathcal Q}_l,\\(n,j)\neq(k,i)}}p_{n,j}\left(\tilde{\mathbf V}\hat{\bm h}_n\hat{\bm h}_n^H\tilde{\mathbf V}^H+\tilde{\mathbf V}\mathbf{\Phi}_n\tilde{\mathbf V}^H\right)+\sigma^2\mathbf I.
\end{aligned}
\end{eqnarray}
Next, we approximate the non-smooth minimum function by the LogSumExp technique. With the LogSumExp, the minimum function is approximated as
\begin{eqnarray}
\begin{aligned}
\min_{k\in\mathcal K} \sum\limits_{i=1}^2\bar r_{k,i}^{\text{lower}}\approx -\alpha \log\left(\frac{1}{K}\sum\limits_{k\in\mathcal K}\exp\left(\frac{ \sum\limits_{i=1}^2\bar r_{k,i}^{\text{lower}}}{-\alpha}\right)\right)
\end{aligned}
\end{eqnarray}
Then (P6) is transformed to
\begin{eqnarray}
\begin{aligned}
\text{(P6.1)}\quad\max_{\bm g_{k,i}}& \quad -\alpha \log\left(\frac{1}{K}\sum\limits_{k\in\mathcal K}\right.\\&\left.\exp\left( \sum\limits_{i=1}^2\log_2\left(\frac{\bm g_{k,i}^H\bar{\mathbf {\Gamma}}_{k,i}^u\bm g_{k,i}}{\bm g_{k,i}^H\bar{\mathbf {\Gamma}}_{k,i}^d\bm g_{k,i}}\right)^{-\frac{1}{\alpha}}\right)\right)\\
\text{s.t.} & \quad \sum\limits_{i=1}^2 \|\bm g_{k,i}\|_2^2\leq P_{\text{max}}^{\text{b}}, \forall k\in \mathcal K.
\end{aligned}\label{eq:1b}
\end{eqnarray}
According to Theorem 1, (P6.1) is an NEPv and can be solved by Algorithm 2.

\subsection{Optimizing Phase-shift Beamforming}
\label{opt_Q2}
With the given decoding order, receive beamforming, and power allocation, the subproblem of (P5) becomes 
\begin{eqnarray}
\begin{aligned}
\text{(P7)}\quad \max_{\bm v}& \quad \min_{k\in\mathcal K} \sum\limits_{i=1}^2\bar r_{k,i}^{\text{lower}}\\
\text{s.t.} & \quad |[\bm v]_n| =1, n=1,\ldots,N,[\bm v]_{N+1}=1\label{eq:p2.3}.
\end{aligned}
\end{eqnarray}
We can rewrite $\bar r_{k,i}^{\text{lower}}$ as (\ref{eq:rlowv}),
\begin{figure*}[!t]
	\vspace*{-8pt}
	\normalsize
	\begin{eqnarray}
	\begin{aligned}
	\bar r_{k,i}^{\text{lower}}
	\\
	=& \log_2\left(1+\frac{p_{k,i}\text{tr}\left(\tilde{\mathbf G}_{k,i}\hat{\bm h}_k^{\dagger}\hat{\bm h}_k^T\tilde{\mathbf G}_{k,i}^H\mathbf V\right)}{p_{k,i}\text{tr}\left(\tilde{\mathbf G}_{k,i}\mathbf{\Phi}_n^{\dagger}\tilde{\mathbf G}_{k,i}^H\mathbf V\right)+\sum\limits_{\substack{(n,j)\in\underline{\mathcal Q}_l,\\(n,j)\neq(k,i)}}p_{n,j}\text{tr}\left(\tilde{\mathbf G}_{k,i}\hat{\bm h}_n^{\dagger}\hat{\bm h}_n^T\tilde{\mathbf G}_{k,i}^H\mathbf V+\tilde{\mathbf G}_{k,i}\mathbf{\Phi}_n^{\dagger}\tilde{\mathbf G}_{k,i}^H\mathbf V\right)+\frac{\text{tr}\left(\tilde{\mathbf G}_{k,i}^H\bm \tilde{\mathbf G}_{k,i}\right)\sigma^2}{(N+1)}}\right).\label{eq:rlowv}
	\end{aligned}
	\end{eqnarray}
	\vspace*{-8pt}
\end{figure*}
where $\tilde{\mathbf G}_{k,i}=\left[[\bm g_{k,i}]_{1}\mathbf I_{(N+1)},\ldots,[\bm g_{k,i}]_{M}\mathbf I_{(N+1)}\right]\in\mathbb C^{(N+1)\times M(N+1)}$.
Then we can turn (P7) into 
\begin{eqnarray}
\begin{aligned}
\text{(P7.1)}\quad \max_{\mathbf V, r }& \quad r\\
\text{s.t.} &\quad\text{(C2)},\text{(C3)},
\\&\quad\text{(C7)}\quad \sum_{i=1}^2 \bar u_{k,i}-\bar d_{k,i}\ge r, \forall k\in\mathcal K,
\end{aligned}
\end{eqnarray}
where $\bar u_{k,i}$ and $\bar d_{k,i}$ are defined as (\ref{eq:ud}).
\begin{figure*}[!t]
	\vspace*{-8pt}
	\normalsize
	\begin{eqnarray}
	\begin{aligned}
	\bar u_{k,i}&\triangleq\log_2\left(\sum\limits_{(n,j)\in\underline{\mathcal Q}_l}p_{n,j}\text{tr}\left(\tilde{\mathbf G}_{k,i}\hat{\bm h}_n^{\dagger}\hat{\bm h}_n^T\tilde{\mathbf G}_{k,i}^H\mathbf V+\tilde{\mathbf G}_{k,i}\mathbf{\Phi}_n^{\dagger}\tilde{\mathbf G}_{k,i}^H\mathbf V\right)+\frac{\sigma^2\text{tr}\left(\tilde{\mathbf G}_{k,i}^H\bm \tilde{\mathbf G}_{k,i}\right)}{(N+1)}\right),\\
	\bar d_{k,i}&\triangleq\log_2\left(p_{k,i}\text{tr}\left(\tilde{\mathbf G}_{k,i}\mathbf{\Phi}_n^{\dagger}\tilde{\mathbf G}_{k,i}^H\mathbf V\right)+\sum\limits_{\substack{(n,j)\in\underline{\mathcal Q}_l,\\(n,j)\neq(k,i)}}p_{n,j}\text{tr}\left(\tilde{\mathbf G}_{k,i}\hat{\bm h}_n^{\dagger}\hat{\bm h}_n^T\tilde{\mathbf G}_{k,i}^H\mathbf V+\tilde{\mathbf G}_{k,i}\mathbf{\Phi}_n^{\dagger}\tilde{\mathbf G}_{k,i}^H\mathbf V\right)
	+\frac{\sigma^2\text{tr}\left(\tilde{\mathbf G}_{k,i}^H\bm \tilde{\mathbf G}_{k,i}\right)}{(N+1)}\right).\label{eq:ud}
	\end{aligned} 
	\end{eqnarray}
	\hrulefill
	\vspace*{-8pt}
\end{figure*}
Like Section \ref{opt_V}, (P7.1) can be approximated as
\begin{eqnarray}
\begin{aligned}
\text{(P7.2)}\quad\max_{\mathbf V, r}& \quad  r -\frac{1}{2\rho_2}\left(\text{tr}\left(\mathbf V\right)-\|\mathbf V^t\|_2\right.\\&\qquad\left.-\text{tr}\left(\bm\alpha_{\text{max}}^t(\bm\alpha_{\text{max}}^t)^H\left(\mathbf V-\mathbf V^t\right)\right)\right)\\
\text{s.t.}&\quad \text{(C2)},(\overline{\text{C3}}),
\\&\quad(\overline{\text{C7}})\quad \sum_{i=1}^2 \bar u_{k,i}-\bar d_{k,i}^t(\mathbf V)\ge r, \forall k\in\mathcal K,
\end{aligned}
\end{eqnarray}
where
\begin{eqnarray}
\begin{aligned}	
\bar d_{k,i}^t(\mathbf V) \triangleq \bar d_{k,i}(\mathbf V^t)+\text{tr}\left(\left(\nabla_{\mathbf V}\bar d_{k,i}(\mathbf V^t)\right)^T\left(\mathbf V-\mathbf V^t\right)\right).
\end{aligned}
\end{eqnarray}
$\nabla_{\mathbf V}\bar d_{k,i}$ is the gradient of function $\bar d_{k,i}$ with respect to $\mathbf V$.
(P7.2) is a convex SDP problem that can be solved by existing CVX.

\section{Numerical results}
\label{Nu}
In this section, we analyze the numerical results of the proposed scheme. The system settings are as follows. The bandwidth $B_s$ is $10$ MHz. The power constraints $P_{\text{max}}$ and $P_{\text{max}}^{\text{b}}$ are $1$ dBm and $30$ dBm, respectively. The complex path gain $\beta_{p}^{rb}$,  $\beta_{p,k}^{sr}$ and  $\beta_{p}^{d}$ follow the Gaussian distribution $\mathcal{CN}(0,1)$. The path delay $\tau_{p}^{rb}$, $\tau_{p,k}^{sr}$ and $\tau_{p,k}^{d}$  follow the uniform distribution $U(0,1/B_s)$. Additionally, the number of MPCs $N_{rb}, N_{sr,k}$, and $N_{d,k}$ varies from $8$ to $16$. Next, we show the numerical results obtained by the proposed optimization scheme.

\begin{figure}[t]
	\centering
	\begin{center}
		\includegraphics[scale=0.5]{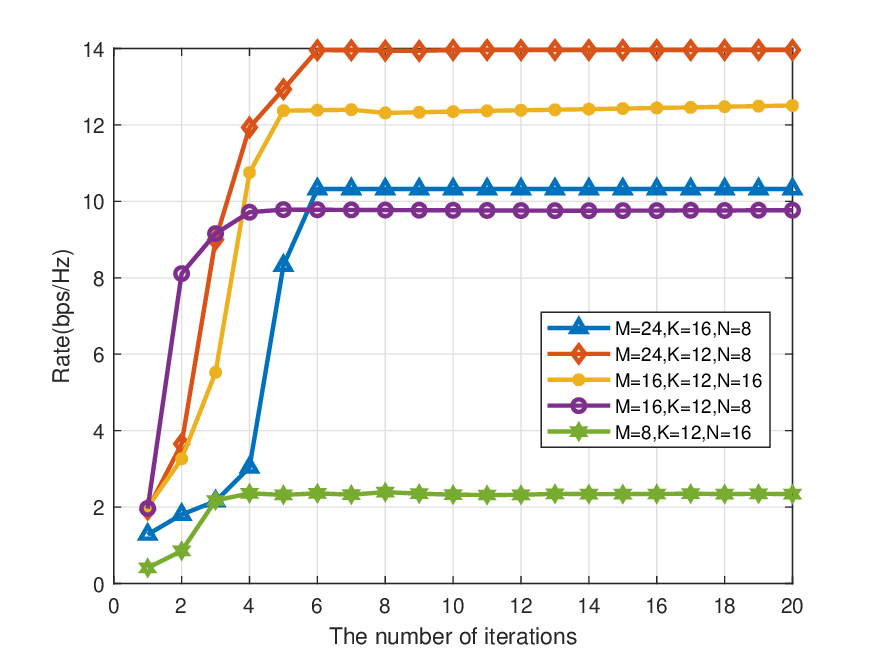}
	\end{center}
	\vspace{-0.3cm}
	\caption{The minimum rate versus the number of iterations. There are SNR$=10$ dB and $L=4$. We consider the CSIT to be known.}
	\label{fig5}
	\vspace{-0.5cm}
\end{figure}
 Figure \ref{fig5} illustrates the convergence behavior of the proposed scheme for different numbers of antennas $N$, IRS reflecting elements $M$, and devices $K$. As depicted in Fig. \ref{fig5}, the proposed algorithm demonstrates convergence for all considered values of $N$, $M$, and $K$. Specifically, when $M=8, K=12, N=16$, the proposed scheme converges after approximately 4 iterations. However, for the case with more antennas, i.e., $M=16, K=12, N=16$, the proposed scheme requires an average of 5 iterations to converge. Similarly, for the case of more antennas,  IRS reflecting elements, and devices, the number of iterations required for the convergence of the scheme increases further. This increase in iteration count can be attributed to the higher number of optimization variables and constraints in the problem, resulting in a more complex optimization process.

The beamforming optimization (P1) is typically relaxed as an SDP problem \cite{RN475, RN473, RN419}. The computational complexity associated with solving the relaxed SDP problem using the interior point method is $\mathcal{O}(KM^{3.5}+M^{2.5}K^2+\sqrt{M}K^3)$ \cite{RN419}. As described in Section \ref{Com}, the computational complexity of the GPI algorithm proposed in this paper to solve (P1) is $\mathcal{O}(KIM^3)$. In theory, the computational complexity of GPI is lower than that of SDP.
Table \ref{tab:my_label} depicts the average running time per iteration for GPI and SDP, with $M=16$ and $L=4$. In both setups, the computation time of SDP significantly exceeds that of GPI.  The numerical results demonstrate the advantage of the proposed method in terms of computational complexity, which is beneficial not only theoretically but also practically. The significant complexity reduction comes from two reasons: 1) By approximating the non-smooth minimum function using a LogSumExp technique, we avoid imposing distinct constraints on the minimum rate in the optimization problem. 2) The use of GPI eliminates the dependence on off-the-shelf solvers like CVX, contributing to computational efficiency.

\begin{table}[h]
	\centering
	\footnotesize
	\renewcommand{\arraystretch}{1.5}
	\setlength{\tabcolsep}{20pt}
	\caption{Average MATLAB CPU time (sec)}
	\label{tab:my_label}
	\scalebox{.8}{
		\begin{tabular}{lll}
			\toprule
			Setup&GPI&SDP \\ \midrule
			$K=12,N=16$&	0.0851&	173.1072\\
			$K=12,N=8$&	0.0305&	125.587\\
			$K=8,N=16$&	0.0573&	97.6686\\
			$K=8,N=8$&	0.0181&	79.114\\
			\bottomrule
		\end{tabular}
	}
\end{table}

\begin{figure}[t]
	\centering
	\subfloat[]{\includegraphics[width=2.9in]{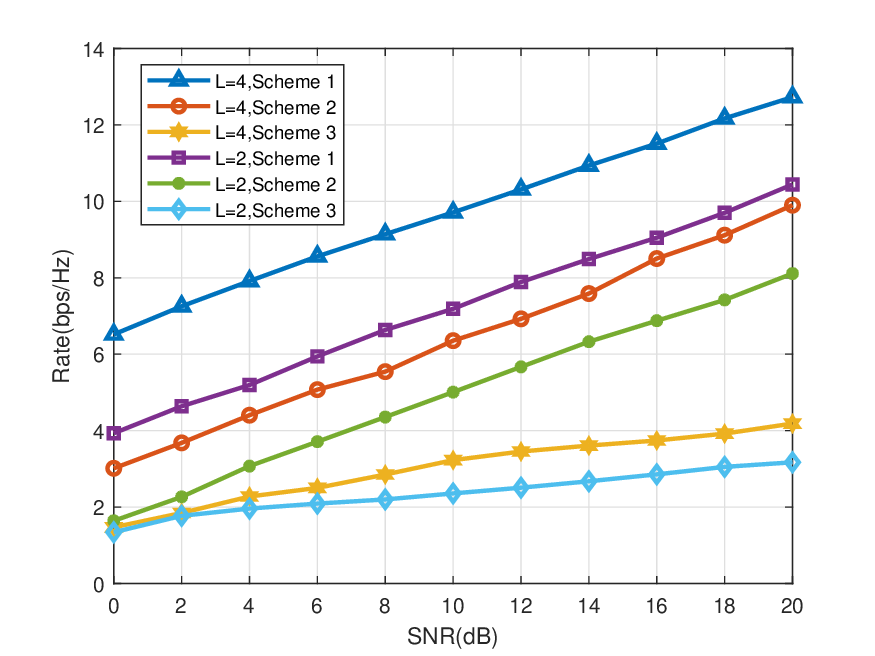}%
		\label{fig:1a}}
	\\
	\subfloat[]{\includegraphics[width=2.9in]{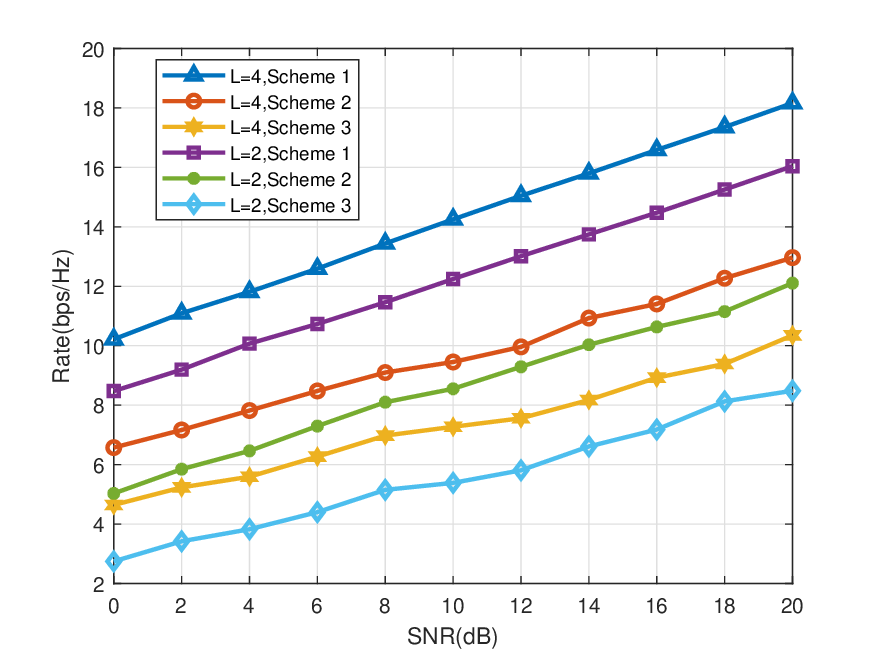}%
		\label{fig:1b}}
	\caption{The minimum rate versus SNR. Scheme 1 is the proposed scheme. Scheme 2 is without IRS-aided scheme and Scheme 3 adopts SDP to optimize the receive beamforming. The setup is as follows: (a). $K=12,M=16,N=8,I=2$; (b). $K=8,M=16,N=16,I=2$. In both cases, we consider the CSIT to be known. }
	\label{fig:1}
	\vspace{-0.5cm}
\end{figure}

To assess the effectiveness of the proposed scheme, we consider the following three schemes: 1) The proposed scheme with known CSIT; 2) The system without IRS-aided; 3) The receive beamforming optimization adopts SDP. Figure \ref{fig:1} shows the minimum rate of the three schemes versus the signal-to-noise ratio (SNR) under different system parameter settings. As $L$ increases from $2$ to $4$, the minimum rate increases for all three schemes. However, compared to other schemes, the proposed scheme exhibits superior performance in enhancing the minimum rate. Even with $L = 2$, it outperforms the other schemes with $L=4$. 
In addition, the numerical results demonstrate that Scheme 2 outperforms Scheme 3, indicating that the benefits derived from the GPI algorithm outweigh those from SDP in an IRS-aided system. There are several key reasons for this observation: 1) GPI only requires a single approximation of (P1), while converting (P1) into a relaxed SDP problem necessitates two approximations. Consequently, the solution obtained from SDP is more prone to errors compared to the solution of (P1) \cite{RN533}; 2) GPI derives the optimal solution from the extreme points, leading to a more stable acquisition of locally optimal solutions.

\begin{figure}[t]
	\centering
	\begin{center}
		\includegraphics[scale=0.5]{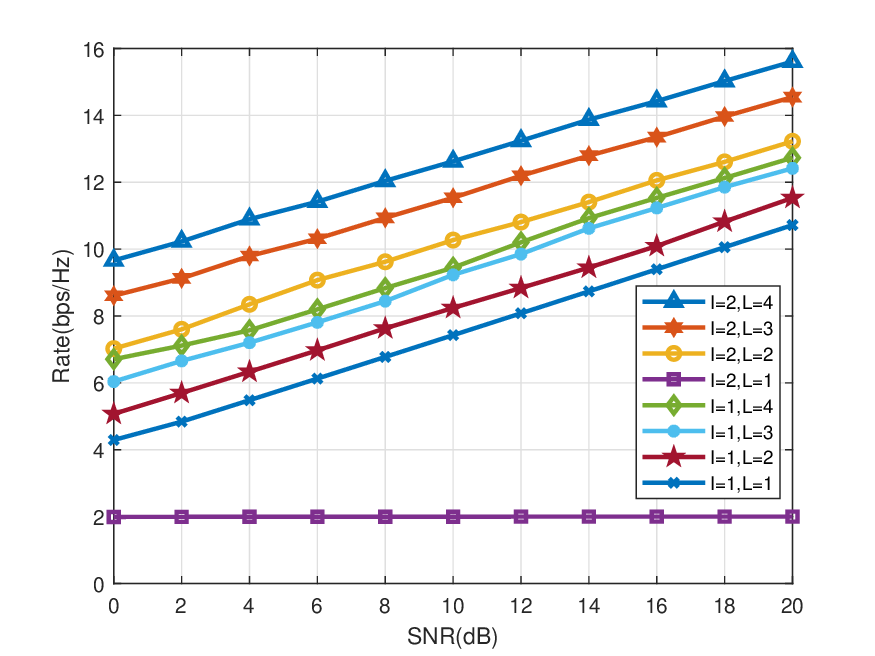}
	\end{center}
	\vspace{-0.3cm}
	\caption{The minimum rate versus SNR. There are $N=M=16$ and $K=12$ with different $I$ and $L$. In this case, the CSIT is known.}
	\label{fig1}
	\vspace{-0.5cm}
\end{figure}

Figure \ref{fig1} shows the minimum rate versus SNR. It is observed that the minimum rate increases as SNR increases. In addition, the minimum rate of devices varies with different $L$ and $I$. When the number of sub-messages for each device $I=1$, the RSMA system reduces to the NOMA system, and the minimum rate increases as the number of groups $L$ increases. When $I=2$ and $L=1$, the rate is initially low and exhibits minimal improvement. This phenomenon can be attributed to the fact that the CSIT is identical for sub-messages originating from the same device, which negatively impacts decoding performance. However, as $L$ increases, the rate for RSMA ($I=2$) is higher than that of the conventional NOMA ($I=1$), demonstrating the effectiveness of RSMA in enhancing system performance. In conclusion, the results illustrate the suitability of SGD for RSMA, emphasizing its role in enhancing system fairness compared to conventional NOMA systems.

\begin{figure}[t]
	\centering
	\begin{center}
		\includegraphics[scale=0.5]{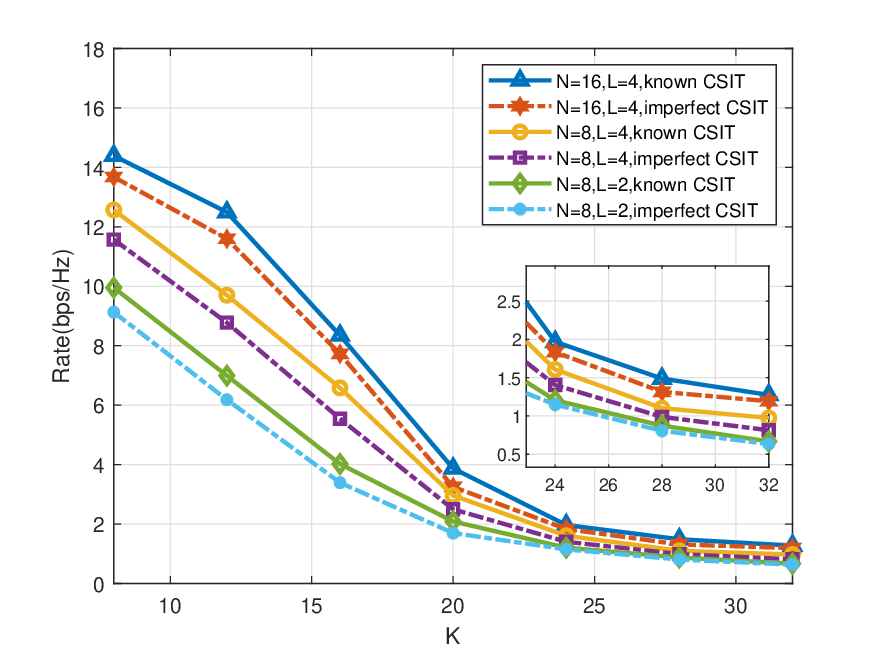}
	\end{center}
	\vspace{-0.2cm}
	\caption{The minimum rate versus the number of devices $K$. There are $M=16$ and SNR$=$$10$ dB with different $N$ and $L$. The numerical simulation considers two cases: known CSIT and imperfect CSIT.}
	\label{fig2}
	\vspace{-0.5cm}
\end{figure}

Figure \ref{fig2} illustrates the minimum rate versus the number of devices $K$. It is obvious that the rate decreases as $K$ increases since the interference becomes higher as $K$ increases. 
Moreover, when considering imperfect CSIT, the minimum rate experiences a further reduction compared to scenarios with known CSIT. This observation demonstrates the detrimental impact of imperfect CSIT on system performance. However, the influence of imperfect CSIT on the minimum rate is relatively modest and diminishes as the number of devices $K$ increases. This trend suggests that the adverse effects of imperfect CSIT can be mitigated to some extent in scenarios with larger device populations.

\begin{figure}[t]
	\centering
	\begin{center}
		\includegraphics[scale=0.5]{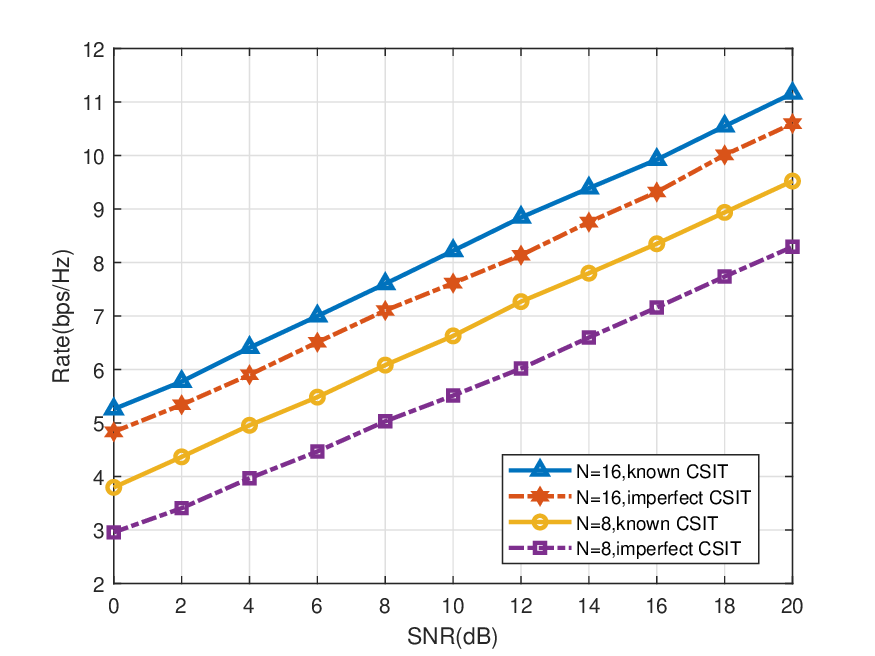}
	\end{center}
	\vspace{-0.3cm}
	\caption{The minimum rate versus SNR. There are $M=16$, $L=4$ and $K=16$ with different $N$. The numerical simulation considers two cases: known CSIT and imperfect CSIT.}
	\label{fig3}
	\vspace{-0.5cm}
\end{figure}

Considering different $N$, Fig. \ref{fig3} describes the performance of the cases with known CSIT and imperfect CSIT. The minimum rate is consistently higher when $N=16$ compared to $N=8$, indicating the beneficial impact of increasing the number of elements in the IRS. Additionally, it is evident that the influence of CSIT errors is more pronounced in the case with $N=8$ compared to that with $N=16$.
Hence, increasing the number of elements in the IRS not only enhances system performance but also improves its robustness against imperfect CSIT.

\begin{figure}[t]
	\centering
	\begin{center}
		\includegraphics[scale=0.5]{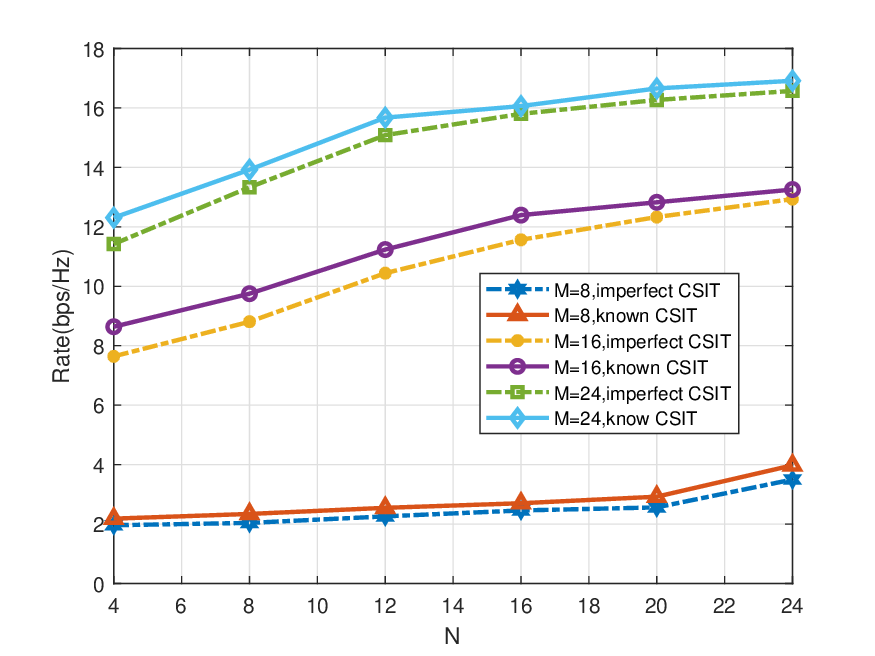}
	\end{center}
	\vspace{-0.3cm}
	\caption{The minimum rate versus $N$. There are SNR$=10$ dB, $L=4$ and $K=12$ with different $M$. The numerical simulation considers two cases: known CSIT and imperfect CSIT.}
	\label{fig4}
	\vspace{-0.5cm}
\end{figure}

Figure \ref{fig4} illustrates the relationship between the minimum rate and the number of reflecting elements $N$ for varying numbers of antennas $M$.
The results indicate that the minimum rate generally increases with both $M$ and $N$. However, the rate increase diminishes as $M$ grows larger. For instance, when $N = 12$, the minimum rate rises from $2.5$ bps/Hz to $11.0$ bps/Hz as $M$ increases from 8 to 16, and from $11.0$ bps/Hz to $15.7$ bps/Hz as $M$ further increases from 16 to 24. This suggests that the rate improvement resulting from additional antennas becomes less significant as $M$ becomes larger.

\section{Conclusion}
\label{Con}
In this work, we proposed an optimization scheme to obtain the minimum rate of devices in the uplink RSMA system with IRS-aided. On the receiving end, we proposed SGD instead of SIC to reduce the decoding complexity and latency. To maximize the minimum rate, we constructed the problem as a max-min optimization problem and adopted the AO algorithm to solve power allocation, beamforming design, and group decoding order, respectively. Specifically, we approximated a non-smooth minimum function using the LogSumExp technique and applied a GPI method to optimize receive beamforming, which exhibited lower complexity compared to the existing method. Additionally, power allocation and phase-shift beamforming design were approximated as convex optimization problems and could be efficiently solved by off-the-shelf solvers. A greedy grouping algorithm was proposed to solve the group decoding order optimization. Furthermore, we extended the proposed scheme to consider scenarios with imperfect CSIT. Numerical results demonstrated that the proposed scheme has superiority in improving the fairness of uplink RSMA. 
In scenarios with imperfect CSIT, the proposed scheme can guarantee the robustness of the system.


\appendices

\section{Proof of Lemma 1}

\label{appendix:l1}

Define the function 
\begin{eqnarray}
\begin{aligned}
&f(\{\bm g_{n,j}\}_{(n,j)\in\mathcal Q})\\&= -\alpha \log\left(\frac{1}{K}\sum\limits_{n\in\mathcal K}\exp\left( \sum\limits_{j=1}^2\log_2\left(\frac{\bm g_{n,j}^H\mathbf {\Gamma}_{n,j}^u\bm g_{n,j}}{\bm g_{n,j}^H\mathbf {\Gamma}_{n,j}^d\bm g_{n,j}}\right)^{-\frac{1}{\alpha}}\right)\right). \nonumber
\end{aligned}
\end{eqnarray}
To find a stationary point, we take the partial derivatives of $f(\{\bm g_{n,j}\}_{(n,j)\in\mathcal Q})$ with respect to $\{\bm g_{k,i}\}_{(k,i)\in\mathcal Q}$ and set it to zero. There is
\begin{eqnarray}
\begin{aligned}
&\frac{\partial f(\{\bm g_{n,j}\}_{(n,j)\in\mathcal Q})}{\partial \bm g_{k,i}}
=\frac{\exp\left( \sum\limits_{j=1}^2\log_2\left(\frac{\bm g_{k,j}^H\mathbf {\Gamma}_{k,j}^u\bm g_{k,j}}{\bm g_{k,j}^H\mathbf {\Gamma}_{k,j}^d\bm g_{k,j}}\right)^{-\frac{1}{\alpha}}\right)}{\sum\limits_{n\in\mathcal K}\exp\left( \sum\limits_{j=1}^2\log_2\left(\frac{\bm g_{n,j}^H\mathbf {\Gamma}_{n,j}^u\bm g_{n,j}}{\bm g_{n,j}^H\mathbf {\Gamma}_{n,j}^d\bm g_{n,j}}\right)^{-\frac{1}{\alpha}}\right)}\\&\qquad\qquad\times \left(\frac{\bm g_{k,i}^H\mathbf {\Gamma}_{k,i}^u}{\bm g_{k,i}^H\mathbf {\Gamma}_{k,i}^u\bm g_{k,i}}-\frac{\bm g_{k,i}^H\mathbf {\Gamma}_{k,i}^d}{\bm g_{k,i}^H\mathbf {\Gamma}_{k,i}^d\bm g_{k,i}}\right).\nonumber
\end{aligned}
\end{eqnarray}
Set the partial derivatives to zero, we have 
\begin{eqnarray}
\begin{aligned}
&\frac{\partial f(\{\bm g_{n,j}\}_{(n,j)\in\mathcal Q})}{\partial \bm g_{k,i}}=\bm 0
\\&\Rightarrow \frac{\exp\left( \sum\limits_{j=1}^2\log_2\left(\frac{\bm g_{k,j}^H\mathbf {\Gamma}_{k,j}^u\bm g_{k,j}}{\bm g_{k,j}^H\mathbf {\Gamma}_{k,j}^d\bm g_{k,j}}\right)^{-\frac{1}{\alpha}}\right)}{\sum\limits_{n\in\mathcal K}\exp\left( \sum\limits_{j=1}^2\log_2\left(\frac{\bm g_{n,j}^H\mathbf {\Gamma}_{n,j}^u\bm g_{n,j}}{\bm g_{n,j}^H\mathbf {\Gamma}_{n,j}^d\bm g_{n,j}}\right)^{-\frac{1}{\alpha}}\right)} \frac{\bm g_{k,i}^H\mathbf {\Gamma}_{k,i}^u}{\bm g_{k,i}^H\mathbf {\Gamma}_{k,i}^u\bm g_{k,i}}\\&=\frac{\exp\left( \sum\limits_{j=1}^2\log_2\left(\frac{\bm g_{k,j}^H\mathbf {\Gamma}_{k,j}^u\bm g_{k,j}}{\bm g_{k,j}^H\mathbf {\Gamma}_{k,j}^d\bm g_{k,j}}\right)^{-\frac{1}{\alpha}}\right)}{\sum\limits_{n\in\mathcal K}\exp\left( \sum\limits_{j=1}^2\log_2\left(\frac{\bm g_{n,j}^H\mathbf {\Gamma}_{n,j}^u\bm g_{n,j}}{\bm g_{n,j}^H\mathbf {\Gamma}_{n,j}^d\bm g_{n,j}}\right)^{-\frac{1}{\alpha}}\right)} \frac{\bm g_{k,i}^H\mathbf {\Gamma}_{k,i}^d}{\bm g_{k,i}^H\mathbf {\Gamma}_{k,i}^d\bm g_{k,i}}.\nonumber
\end{aligned}
\end{eqnarray}
Then there is 
\begin{eqnarray}
\begin{aligned}
&\mathbf \Omega_{k,i}^{-1}\left(\{\bm g_{n,j}\}_{(n,j)\in\mathcal Q}\right)\mathbf \Psi_{k,i}\left(\{\bm g_{n,j}\}_{(n,j)\in\mathcal Q}\right)\bm g_{k,i}\\&\qquad=\lambda (\{\bm g_{n,j}\}_{(n,j)\in\mathcal Q})\bm g_{k,i}.
\end{aligned}
\end{eqnarray}
The proof is complete.
\section{Proof of Lemma 2}

\label{appendix:l2}

Since $\mathbf A$ is a $M(N+1)\times M(N+1)$ Hermitian matrix, it is diagonalizable and has $M(N+1)$ linearly independent eigenvectors denoted as $\bm u_1,\bm u_2\ldots,\bm u_{M(N+1)}$. With loss of generality, let's assume that $\bm u_1,\bm u_2\ldots,\bm u_{M(N+1)}$ correspond to eigenvalues of $ \lambda_1>\lambda_2\ge\ldots\ge\lambda_{M(N+1)}$. Then, the initial approximation $\bm u_0$ can be written as a linear combination of these eigenvectors:
\begin{eqnarray}
\begin{aligned}
\bm u_0=\alpha_1\bm u_1+\alpha_2\bm u_2+\ldots+\alpha_{M(N+1)}\bm u_{M(N+1)},
\label{eq:l2}
\end{aligned}
\end{eqnarray}
with $\alpha_1\ne 0$. (If $\alpha_1= 0$, the power method may not converge, and $\bm u_0$ needs to be reselected.)  Now, multiplying both sides of (\ref{eq:l2}) by $\mathbf A$ has 
\begin{eqnarray}
\begin{aligned}
\mathbf A\bm u_0&=\mathbf A(\alpha_1\bm u_1+\ldots+\alpha_{M(N+1)}\bm u_{M(N+1)})\\&=\alpha_1(\mathbf A\bm u_1)+\ldots+\alpha_{M(N+1)}(\mathbf A\bm u_{M(N+1)})\\&=\alpha_1(\lambda_1\bm u_1)+\ldots+\alpha_{M(N+1)}(\lambda_{M(N+1)}\bm u_{M(N+1)}).\nonumber
\end{aligned}
\end{eqnarray}
Therefore, repeat this operation $t'$ times and there is
\begin{eqnarray}
\begin{aligned}
\mathbf A^{t'}\bm u_0&=\alpha_1(\lambda_1^{t'}\bm u_1)+\alpha_2(\lambda_2^{t'}\bm u_2)+\ldots\\&\qquad\qquad+\alpha_{M(N+1)}(\lambda_{M(N+1)}^{t'}\bm u_{M(N+1)})\\&=\lambda_1^{t'}\left(\alpha_1\bm u_1+\alpha_2(\frac{\lambda_2}{\lambda_1})^{t'}\bm u_2+\ldots\right.\\&\qquad\qquad\left.+\alpha_{M(N+1)}(\frac{\lambda_{M(N+1)}}{\lambda_1})^{t'}\bm u_{M(N+1)}\right).\nonumber
\end{aligned}
\end{eqnarray}
Since $ \lambda_1>\lambda_2\ge\ldots\ge\lambda_{M(N+1)}$, it implies that
\begin{eqnarray}
\begin{aligned}
\lim_{{t'}\rightarrow \infty}\mathbf A^{t'}\bm u_0&=\lim_{{t'}\rightarrow \infty}\lambda_1^{t'}\left(\alpha_1\bm u_1+\alpha_2(\frac{\lambda_2}{\lambda_1})^{t'}\bm u_2+\ldots\right.\\&\qquad\left.+\alpha_{M(N+1)}(\frac{\lambda_{M(N+1)}}{\lambda_1})^{t'}\bm u_{M(N+1)}\right)\\&=\lambda_1^{t'}\alpha_1\bm u_1. 
\end{aligned}
\end{eqnarray}
Therefore, Lemma 2 holds.

\section{Proof of Theorem 2}
\label{appendix:t3}
For convenience, we use $\mathbf \Omega_{k,i}(t-1)$ and $\mathbf \Psi_{k,i}(t-1)$ instead of $\mathbf \Omega_{k,i}(\{\bm g_{k,i}^s(t-1)\}_{(k,i)\in\mathcal Q})$ and $\mathbf \Psi_{k,i}(\{\bm g_{k,i}^s(t-1)\}_{(k,i)\in\mathcal Q})$, respectively.
First, according to (\ref{eq:phi}), $\mathbf A_{k,i}(t)$ can be expressed as
\begin{eqnarray}
\begin{aligned}
\mathbf A_{k,i}(t)
=\frac{(\bm g_{k,i}^s(t-1))^H\mathbf {\Gamma}_{k,i}^d\bm g_{k,i}^s(t-1)}{(\bm g_{k,i}^s(t-1))^H\mathbf {\Gamma}_{k,i}^u\bm g_{k,i}^s(t-1)}
({\mathbf {\Gamma}_{k,i}^d})^{-1}\mathbf {\Gamma}_{k,i}^u\lambda (t-1).\nonumber
\end{aligned}
\end{eqnarray}
We first prove that
\begin{eqnarray}
\begin{aligned}
\bm g_{k,i}^s(t+t'-1)&=\frac{\mathbf A_{k,i}(t+t'-1)\bm g_{k,i}^s(t+t'-2)}{\|\mathbf A_{k,i}(t+t'-1)\bm g_{k,i}^s(t+t'-2)\|_2}\\&=\frac{\left(\left({\mathbf {\Gamma}_{k,i}^d}\right)^{-1}\mathbf {\Gamma}_{k,i}^u\right)^{t'}\bm g_{k,i}^s(t-1)}{\left\|\left(\left({\mathbf {\Gamma}_{k,i}^d}\right)^{-1}\mathbf {\Gamma}_{k,i}^u\right)^{t'}\bm g_{k,i}^s(t-1)\right\|_2}.
\label{eq:t2p}
\end{aligned}
\end{eqnarray} 
When $t'=1$, there is
\begin{eqnarray}
\begin{aligned}
\bm g_{k,i}^s(t)=
\frac{\mathbf A_{k,i}(t)\bm g_{k,i}^s(t-1)}{\|\mathbf A_{k,i}(t)\bm g_{k,i}^s(t-1)\|_2}=\frac{({\mathbf {\Gamma}_{k,i}^d})^{-1}\mathbf {\Gamma}_{k,i}^u\bm g_{k,i}^s(t-1)}{\|({\mathbf {\Gamma}_{k,i}^d})^{-1}\mathbf {\Gamma}_{k,i}^u\bm g_{k,i}^s(t-1)\|_2}.\nonumber
\end{aligned}
\end{eqnarray}
Eq.(\ref{eq:t2p}) holds.
When $t'>1$, we assume that (\ref{eq:t2p}) holds for $t'-1$, i.e.,
\begin{eqnarray}
\begin{aligned}
\bm g_{k,i}^s(t+t'-2)=\frac{\left(\left({\mathbf {\Gamma}_{k,i}^d}\right)^{-1}\mathbf {\Gamma}_{k,i}^u\right)^{t'-1}\bm g_{k,i}^s(t-1)}{\left\|\left(\left({\mathbf {\Gamma}_{k,i}^d}\right)^{-1}\mathbf {\Gamma}_{k,i}^u\right)^{t'-1}\bm g_{k,i}^s(t-1)\right\|_2}.\nonumber
\end{aligned}
\end{eqnarray} 
Then, there is
\begin{eqnarray}
\begin{aligned}
\bm g_{k,i}^s(t+t'-1)&=\frac{\mathbf A_{k,i}(t+t'-1)\bm g_{k,i}^s(t+t'-2)}{\|\mathbf A_{k,i}(t+t'-1)\bm g_{k,i}^s(t+t'-2)\|_2}
\\&=\frac{\left(\left({\mathbf {\Gamma}_{k,i}^d}\right)^{-1}\mathbf {\Gamma}_{k,i}^u\right)^{t'}\bm g_{k,i}^s(t-1)}{\left\|\left(\left({\mathbf {\Gamma}_{k,i}^d}\right)^{-1}\mathbf {\Gamma}_{k,i}^u\right)^{t'}\bm g_{k,i}^s(t-1)\right\|_2}.
\end{aligned}
\end{eqnarray}
Therefore, (\ref{eq:t2p}) holds for $t'>0$.

Next, we can denote $\mathbf A_{k,i}^{t'}(t)$ as 
\begin{eqnarray}
\begin{aligned}
\mathbf A_{k,i}^{t'}(t)=&\left(\frac{(\bm g_{k,i}^s(t-1))^H\mathbf {\Gamma}_{k,i}^d\bm g_{k,i}^s(t-1)}{(\bm g_{k,i}^s(t-1))^H\mathbf {\Gamma}_{k,i}^u\bm g_{k,i}^s(t-1)}\lambda (t-1)\right)^{t'}
\\&\times\left(\left({\mathbf {\Gamma}_{k,i}^d}\right)^{-1}\mathbf {\Gamma}_{k,i}^u\right)^{t'}.
\end{aligned}
\end{eqnarray} 
Then the left side of (\ref{eq:t2}) is
\begin{eqnarray}
\begin{aligned}
\frac{\mathbf A_{k,i}^{t'}(t)\bm g_{k,i}^s(t-1)}{\|\mathbf A_{k,i}^{t'}(t)\bm g_{k,i}^s(t-1)\|_2}&=\frac{\left(\left({\mathbf {\Gamma}_{k,i}^d}\right)^{-1}\mathbf {\Gamma}_{k,i}^u\right)^{t'}\bm g_{k,i}^s(t-1)}{\left\|\left(\left({\mathbf {\Gamma}_{k,i}^d}\right)^{-1}\mathbf {\Gamma}_{k,i}^u\right)^{t'}\bm g_{k,i}^s(t-1)\right\|_2}\\&=\bm g_{k,i}^s(t+t'-1).\nonumber
\end{aligned}
\end{eqnarray} 
The proof is complete.

\section{Proof of Theorem 3}
\label{appendix:t1}
Define the the objective function $f(\mathbf V)$ as follows
\begin{eqnarray}
\begin{aligned}
f(\mathbf V)=r=\min\limits_{(k,i)\in \mathcal Q} u_{k,i}(\mathbf V)-d_{k,i}^t(\mathbf V).
\end{aligned}
\end{eqnarray}  
Assume that $\mathbf V^*$ is the optimal solution of (P2.1). Then we have $f(\mathbf V)\le f(\mathbf V^*)$ for all $\mathbf V$ which satisfy $\text{tr}(\mathbf V)-\|\mathbf V\|_2=0$. Let $g(\mathbf V,\rho_s)$ and $\mathbf V^s$ denote the objective function and the optimal solution of (P2.2), respectively. With penalty factor $\rho_s$, there is 
\begin{eqnarray}
\begin{aligned}
g(\mathbf V^s,\rho_s)\ge g(\mathbf V^*,\rho_s),
\end{aligned}
\end{eqnarray}
which implies
\begin{eqnarray}
\begin{aligned}
f(\mathbf V^s)-&\frac{1}{2\rho_s}\left(\text{tr}(\mathbf V^s)-\|\mathbf V^s\|_2\right)\ge \\&f(\mathbf V^*)-\frac{1}{2\rho_s}\left(\text{tr}(\mathbf V^*)-\|\mathbf V^*\|_2\right).
\end{aligned}
\end{eqnarray}
Since $\mathbf V^*$ is the optimal solution of (P2.1), the rank-one constraint must be satisfied, $\text{tr}(\mathbf V^*)-\|\mathbf V^*\|_2=0$. The above inequality is written as
\begin{eqnarray}
\begin{aligned}
&f(\mathbf V^s)-\frac{1}{2\rho_s}\left(\text{tr}(\mathbf V^s)-\|\mathbf V^s\|_2\right)\ge f(\mathbf V^*)\label{eq:ap2}\\\Rightarrow&\text{tr}(\mathbf V^s)-\|\mathbf V^s\|_2\le 2\rho_s\left(f(\mathbf V^s)-f(\mathbf V^*)\right).\label{eq:ap1}
\end{aligned}
\end{eqnarray}
Suppose $\bar{\mathbf V}$ is a limit point of sequence $\{{\mathbf V}^s\}$ and exist an infinite subsequence $\mathcal S$ such that $\lim_{s\in \mathcal S} \mathbf V^s = \bar{\mathbf V}$. By taking the limit as $s\to \infty, s\in \mathcal S$ on both side of (\ref{eq:ap1}), there is 
\begin{eqnarray} 
\begin{aligned}
\text{tr}(\bar{\mathbf V})-\|\bar{\mathbf V}\|_2&=\lim_{s\in \mathcal S}\left(\text{tr}(\mathbf V^s)-\|\mathbf V^s\|_2\right)\\&\le \lim_{s\in \mathcal S}2\rho_s\left(f(\mathbf V^*)-f(\mathbf V^s)\right)\stackrel{\rho_s\to 0}{=}0 .
\end{aligned}
\end{eqnarray}
where the left side holds due to the continuity of function $\text{tr}({\mathbf V})-\|{\mathbf V}\|_2$. In a result, there is $\text{tr}(\bar{\mathbf V})-\|\bar{\mathbf V}\|_2=0$. So $\bar{\mathbf V}$ is feasible for (P2.1). By taking the limit as $s\to \infty, s\in \mathcal S$ on (\ref{eq:ap2}), we have 
\begin{align}
f(\bar{\mathbf V})\ge f(\bar{\mathbf V})-\lim_{s\in \mathcal S}\frac{1}{2\rho_s}\left(\text{tr}(\mathbf V^s)-\|\mathbf V^s\|_2\right)\ge f(\mathbf V^*),
\end{align}
where $\rho_s$ and $\text{tr}(\mathbf V^s)-\|\mathbf V^s\|_2$ are non-negative.
Therefore, $\bar{\mathbf V}$ is a set of feasible points whose objective value is no less than that of the optimal solution $\mathbf V^*$. It is obvious that $\bar{\mathbf V}$ is also an optimal solution for (P2.1), which completes the proof.
\ifCLASSOPTIONcaptionsoff
\newpage
\fi
\bibliographystyle{IEEEtran}
\bibliography{main}

\vfill

\end{document}